\documentclass[aps,rmp,floats,twocolumn,showkeys,showpacs,reprint,superscriptaddress]{revtex4-1}

\usepackage[latin1]{inputenc}
\usepackage{amssymb,amsmath,amsfonts}
\usepackage{verbatim}
\usepackage{graphicx}
\usepackage{subfig}
\usepackage{xcolor}
\usepackage{colortbl}
\usepackage{natbib}
\usepackage[section]{placeins}
\usepackage{soul}

\definecolor{MyGreen}{rgb}{0,0.8,0.2}
\definecolor{MyRed}{rgb}{1.0,0.3,0.2}

\definecolor{MyOrange}{rgb}{1.0, 0.48, 0.0}
\definecolor{Enzo}{rgb}{0.00, 0.79, 0.86}

\begin{document}
\title{Assessment of urban ecosystem resilience using the efficiency
  of hybrid social-physical complex networks}

\author{D. Asprone}
\affiliation{Department of Structures for Engineering and Architecture, University of Naples ``Federico II'', Naples, Italy}
\author{M. Cavallaro}
\affiliation{School of Mathematical Sciences, Queen Mary University of London, London, UK}
\author{V. Latora}
\affiliation{School of Mathematical Sciences, Queen Mary University of London, London, UK}
\affiliation{Dipartimento di Fisica ed Astronomia, Universit\`a di Catania and INFN, I-95123 Catania, Italy}
\author{G. Manfredi}
\affiliation{Department of Structures for Engineering and Architecture, University of Naples ``Federico II'', Naples, Italy}
\author{V. Nicosia}
\affiliation{School of Mathematical Sciences, Queen Mary University of London, London, UK}
\date{\today}


\begin{abstract}
One of the most important tasks of urban and hazard planning is to
mitigate the damages and minimize the costs of the recovery process
after catastrophic events. The rapidity and the efficiency of the
recovery process are commonly referred to as resilience. Despite the
problem of resilience quantification has received a lot of attention,
a mathematical definition of the resilience of an urban community,
which takes into account the social aspects of a urban environment,
has not yet been identified. In this paper we provide and test a
methodology for the assessment of urban resilience to catastrophic
events which aims at bridging the gap between the engineering and the
ecosystem approaches to resilience. We propose to model a urban system
by means of different hybrid social-physical complex networks,
obtained by enriching the urban street network with additional
information about the social and physical constituents of a city,
namely citizens, residential buildings and services. Then, we
introduce a class of efficiency measures on these hybrid networks,
inspired by the definition of global efficiency given in complex
network theory, and we show that these measures can be effectively
used to quantify the resilience of a urban system, by comparing their
respective values before and after a catastrophic event and during the
reconstruction process. As a case study, we consider simulated
earthquakes in the city of Acerra, Italy, and we use these efficiency
measures to compare the ability of different reconstruction strategies
in restoring the original performance of the urban system.
\end{abstract}

\keywords{resilience, risk, street networks, complex networks}

\maketitle

\section{Introduction}

Today more than 50\% of humans around the world live in urbanised
areas~\citep{UNPopulationDevel} and this percentage is expected to
increase by 2050 up to 86\% in developed countries and up to 64\% in
developing countries~\citep{UNPopulationProspect}. Consequently, the
estimation of the resilience of urban systems against natural and
human-induced risks, together with the implementation of successful
strategies aiming at resilience strengthening and risk mitigation,
represents a urgent challenge for the scientific community, with
potential impact on billions of human lives.

Cities are among the most complicated and beautiful human artefacts,
being the result of the intricate interaction of several constraints
and processes which determine, reshape and refine the urban structure
at all scales. One of the most relevant of such constraints is
geographical: cities are embedded in a geometrical space, and their
structure and development is heavily and undoubtedly affected ---and to
some extent driven--- by the morphology of the surrounding
environment. Then, there is a technological component, consisting of
all the buildings and services provided by the city itself, together
with the infrastructures that physically link them together: as a
matter of fact, urban areas have become more attractive than rural
settlements only because they guarantee a faster and more efficient
access to a wider set of facilities, services and
opportunities. Finally, and more importantly, there is the human
aspect, i.e. all the citizens that reside, move and work within this
physical frame, together with all the social, economical, cultural and
historical processes that contribute to the evolution of an urban area
over the centuries. Taking into account all these factors, we easily
realise that a city can be effectively considered a \textit{complex
  system}, i.e., according to one of the most general definition
available, a system consisting of many interacting units with the
ability to generate a non-trivial, collective behaviour through the
combination of simple mechanisms acting at a local
scale~\citep{Bar-Yam}.

In the last few decades there has been an increasing interest for the
quantitative study of urban systems from a complex systems
perspective, and several methods and metrics have been proposed to
measure the structural properties of cities and to quantify their
evolution over
time~\citep{Wilson1970,Batty1994,Batty2005Cities,Makse1995,Salingaros2005,Marshall2006,
  Bettencourt2007}, with particular attention to the study of street
patterns and transportation
networks~\citep{Cardillo2006,Crucitti2006PhyRevE,Scellato2006,Bettencourt2010,Porta2011,
  Strano2012}.

Despite recent literature has pointed out the necessity to introduce
ad-hoc metrics to quantify the resilience of a city against
shocks~\citep{Dalziell2004}, to date a unique and universally accepted
definition of urban resilience is still missing. So far, the efforts
to provide an operational measure to assess resilience of urban
systems against shocks and disasters have been inspired by two main
philosophies, namely the \emph{engineering vision} and the
\emph{ecosystem vision}, which are fundamentally different in the
spirit and, to some extent, diverging~\citep{Lorenz2010}.

In the engineering vision, the resilience of a city or a metropolitan
area depends on the capability of all the physical components of the
system, including buildings and transportation infrastructures, to
absorb the damages due to an external shock and to quickly restore
their state before the shock~\citep{O'Rourke2007, Reed2009,
  Bruneau2003, Pimm1984, Opricovic2002}.
%
%

In the ecosystem approach, instead, resilience is defined as the
capability of the whole urban system to recover the full set of
functionalities and services that existed before the shock, even
without returning exactly to the state before the
shock~\citep{Holling1973, Holling1986, Holling2001, Holling2002,
  Folke2006, Kovalenko2012}.  This implies that, due to the
restoration of the infrastructures and services damaged by the shock,
the city might evolve in something slightly different from what it was
before, while keeping its identity in a broader sense.
On the one
hand, the engineering approach
explicitly requires that each single component of the system, due to the
restoration process, reaches a new state in which its performance is
not worse than one it had before the shock occurred.  On the other
hand, the ecosystem approach to resilience provides a more general
framework to understand the recovery of complex
systems~\citep{Holling1996}. However, this approach requires that the
metrics employed to quantify resilience should be able to capture the
performance of the system \emph{as a whole}, which is usually much
more than the algebraic sum of the performances of its single
components.  Determining which of the two approaches is more
appropriate for the quantification of urban resilience is still a
matter of active debate.

In this paper we engage with this debate by presenting a consistent
framework for the quantification of urban systems resilience, which
interpolates between the engineering and the ecosystem approach,
allowing a quantitative estimation of system resilience while
focusing, at the same time, on the performance of the urban system as
a whole. This framework is based on the representation of cities as
complex networks. In particular, we make use of Hybrid
Social-Physical Networks (HSPNs), which provide a compact
representation of the geographical, technological and social aspects
of a city. Then, we propose a set of network efficiency indexes to
quantify the performance of a HSPNs.
We show that the damage inflicted
to a city by a shock can be easily quantified as the ratio between the
efficiency of the corresponding HSPNs after and before the shock.
Similarly, the effectiveness of different reconstruction
strategies is compared by computing the evolution of the network
efficiency of the HSPNs obtained using each strategy.

As a case study we considered the city of Acerra (Italy).  We first
constructed three HSPNs of Acerra in the pre-shock
configuration. Then, we quantified the damage inflicted to the
original networks by simulating earthquakes of increasing intensity.
Finally, we compared the performance of six different reconstruction
strategies in restoring the original HSPN efficiency.
The results suggest that the weakness of the urban ecosystem of Acerra
is due to an high-risk historical centre and that reconstruction
strategies which allocate a large portion of displaced people in few
distant points can not completely restore the pre-shock efficiency.

The paper is organised as follows. In Section~\ref{sec:related} we
briefly discuss the state of the art about urban resilience
quantification, focusing on the theoretical contributions to the field
recently provided by researchers in complexity and network science. In
Section~\ref{sec:methodology} we define three Hybrid Social-Physical
Networks associated to a city and we propose a set of efficiency
metrics to quantify their performance. In
Section~\ref{sec:hspn_resilience} we define a measure to assess the
resilience of a urban system based on the efficiency of the associated
HSPNs, and we review a few reconstruction strategies typically
employed in the aftermath of a disaster. In
Section~\ref{sec:case_study} we report the results of the proposed
methodology on a case study, based on the simulation of earthquakes of
increasing intensity in the city of Acerra (Italy), and we discuss the
ability of different reconstruction strategies to recover the original
performance of the city. Finally, in Section~\ref{sec:conclusions} we
report a discussion of the results and suggest some potential future
extensions of this work.

\section{Motivation and related works}
\label{sec:related}

The resilience of infrastructure and transportation systems against
natural and human-induced disasters has been largely investigated in
the literature and several methods to quantify resilience have been
proposed, analysed and discussed. A comprehensive review of the
subject is beyond the scope of the present work.  However, we believe
that some recent contributions have proposed interesting concepts
which could be potentially employed to quantify the resilience of
urban systems.
For instance, \citet{Reed2009} proposed a method
to characterize the behavior of networked infrastructures prone to
natural hazards, considering the interdependency of the
system, with applications on power and telecommunication networks.
\citet{Ouyang2012} proposed a multi-stage framework to analyse
power transmission grids resilience and have identified, for each
stage, a series of resilience-based improvement strategies.
\citet{Maliszewski2012} hypothesized that the resilience of power
distribution systems depends on two main factors, i.e., the
environment where the network operates, and the priority policy
employed during restoration.
\citet{Attoh-Okine2009} proposed a resilience index for urban
infrastructures based on the Belief Function framework, while
\citet{Li2007} introduced a resilience index defined as the ratio
between the failure probability and the recovery probability.
\citet{Omer2009} focused on the resilience of telecommunication cable
systems, defining it as the ratio between the amount of information
transmitted after a disruption and the amount carried before the
shock occurred. The reliability of infrastructure networks prone to
natural hazards has been largely discussed in~\citep{Pinto2006}, and
several different methods based on connectivity and flow were recently
 analysed~\citep{Li2002,Rojo2011,Cavalieri2012,Franchin2013}.

In a recent paper, \citet{Tamvakis2013} provided a comparative
review of several methods for resilience quantification, pointing out
that most of the existing approaches might actually have quite limited
applicability, due to the fact that these methods usually rely on some
ad-hoc assumptions and often focus on specific subsystems, like
telecommunication or power distribution networks. As a result,
although some of the concepts and methodologies are interesting and
potentially powerful, they cannot be straightforwardly extended to
quantify the resilience of a urban system as a whole, which normally
consist of several interconnected and interdependent subsystems.

In the last decade or so, important contributions to the problem of
measuring the robustness of a system and quantifying its resilience to
attacks and failures have come from the analytical study of complex
networks. Complex network theory has proven to be a robust theoretical
framework to study the topology of networked systems and has largely
been employed for the characterization of a variety of phenomena
occurring in systems composed by interconnected units, including many
biological, technological and social
networks~\citep{Strogatz2001,Newman2003rev,Boccaletti2006}. Recently,
complex network theory has also been successfully employed to quantify
and model the topological aspects of spatial networks in
general~\citep{Barthelemy2011} and of street networks in
particular~\citep{Crucitti2006PhyRevE,Strano2012}.  The complex
network approach to resilience is based on the analysis of an
extremely simplified model ---a graph---, representing the elementary
components of the the original system and the relations among
them. The main assumption is that such a network model, despite
discarding some specific details, is nevertheless able to capture the
fundamental properties of the original system. This approach has been
successfully employed to study the robustness and resilience of
complex transportation networks, information networks and power
distribution systems~\citep{Albert2000, Callaway2000, Cohen2000,
  Paul2004} and has recently been extended to the case of multi-layer
and interdependent networks~\citep{Berche2009,Buldyrev2010,
  Satumtira2010,Vespignani2010,Gao2011, Gao2011a, Huang2013}.

As a matter of fact, the functioning of the infrastructures and
services of a urban area heavily rely on the existence of an
underlying road network, and the efficiency of a city as a whole
undoubtedly depends on the topological properties of its street
pattern. In this respect, the street network is one of the most
important aspects of a city, since its structure is intimately
connected with the reachability of services and facilities and
therefore with the overall quality of life perceived by the citizens.
The quantitative analysis of urban street networks has shown that
their topologies have complex structural
properties~\citep{Cardillo2006,Scellato2006,Porta2011,
  Barthelemy2011}, and recent works seem to
confirm that the street network plays a central role in shaping the
evolution of an urban area
~\citep{Southworth2003,Batty2005Cities,Marshall2006,Bettencourt2007,Strano2012}.
Consequently, it would be tempting to define the resilience of a urban
system in terms of the resilience of its street network.

However, a urban system is indeed the result of the intricate
combination of several technological and social processes, and all its
richness and complexity cannot be fully captured by the analysis of
the underlying street network alone. A meaningful assessment of urban
resilience to shocks should be focused on the real impact of the shock
on the efficiency of a urban system \textit{as perceived by the
  citizens}, and should therefore take into account other factors that
concur to the perceived post-shock lack of performance, including
population density, location of facilities, services availability and
relocation strategies.

In the next Section we introduce a simplified representation of urban
systems based on hybrid complex networks, which provides a consistent
framework to integrate the structure of the urban street network and
information about inhabitants, buildings and services.

\section{Modelling urban systems by means of Hybrid Social-Physical networks}
\label{sec:methodology}

The methodology for the assessment of urban resilience to shocks that
we present here aims at bridging the gap between the engineering and
the ecosystem approach to resilience. On the one hand, we identify a
set of measures which allow to quantify, from an engineering point of
view, the ability of a urban system to return to its pre-shock
performance after a disaster. On the other hand, in the same spirit of
the ecosystem vision, our measures are able to quantify the resilience
of urban system even when, due to the post-shock restoration process,
the urban system has attained a different configuration.

Here we first review some standard metrics for complex street networks
analysis, and we then introduce Hybrid Social-Physical networks,
together with metrics to quantify their overall efficiency.

\subsection{Networks of urban street patterns}
\label{sec:streetpattern}

Generally, networks can conveniently be described by means of graphs
consisting of a set of points $\mathcal{N}$, called nodes or vertices,
and by a set $\mathcal{V}$ of edges connecting pairs of points.  A
graph with $N=|\mathcal{N}|$ nodes ($\mathcal{N}=\{n_1,n_2,n_3, \ldots
n_N\}$) and $K=|\mathcal{V}|$ edges ($\mathcal{V}=\{v_1,v_2,v_3,
\ldots v_k\}$) can be represented by giving its adjacency matrix,
i.e. the $N \times N$ matrix $A=\{a_{ij}\}$ whose entry $a_ {ij}$ is
equal to $1$ if there is an edge connecting node $i$ and node $j$,
while $a_{ij}=0$ otherwise.  It is also possible to assign a weight or
a length $l_{ij}$ to each edge linking nodes $i$ and $j$ in a graph,
thus defining a weighted adjacency matrix~$\{ l_{ij}\}$.

Spatial networks are a special class of complex networks whose nodes
are embedded in a space associated with a metric. Typical examples of
spatial networks include electric power grids~\citep{Kinney2005EPJB}
and transportation systems including rivers, trade routes and street
networks~\citep{Pitts1965TheProfessional,Crucitti2006PhyRevE,Strano2012}.
In the case of street networks, each crossing is represented by a node
while edges represent street segments, so that two nodes are connected
by an edge if the corresponding crossings are adjacent to the same
segment of road.  Given a city, in the following we denote by
$\mathbb{G}\mathcal{(N,V)}$ the graph representing the urban street
pattern, where $\mathcal{N}$ is set of street junctions and
$\mathcal{V}$ is the set of street segments.  Street networks are
naturally embedded in a two-dimensional Euclidean space, whose metric
is the usual Euclidean distance, so that the lengths $l_{ij}$ of the
edges satisfy the triangular equality~\citep{Barthelemy2011}.

In transportation and communication networks it is usually important
to know how to move or send an information from a node $i$ to another
node $j$. An alternate sequence of nodes and edges that starts from
$i$ and ends in $j$ is called a \textit{walk} from $i$ to $j$. If
there exists a walk between node $i$ and node $j$, we say that $i$ and
$j$ are \textit{connected}. A maximal set of nodes which are mutually
connected to each other is called a \textit{component} of the
graph. If not all the pairs of nodes in the graph are connected, then
the graph is composed by more than one \textit{component}. Each walk
is associated to a cost, that is the sum of the lengths of the edges
involved in the walk. If each node of the walk is traversed only once,
then the walk is called a \textit{path}.  The path from i to j having
minimal length is called shortest path and its length is denoted by
$d_{ij}$.  If two nodes are not linked by any walk, then $d_{ij}$ is
set to $\infty$, and the two nodes are said to be
\textit{disconnected}.

A measure of the typical separation between nodes in the graph is the
\textit{characteristic path length} $L$, that is the mean value of the
length of the shortest paths between all the possible pairs of nodes:
\begin{equation}
	L = \frac{1}{N(N-1)}\sum_{i,j\in \mathcal{N}, i\neq j}d_{ij}.
\label{fig:char_path_lenth}
\end{equation}
In general, the lower the characteristic path length, the better the
communication between any pair of nodes chosen at random.
Using the characteristic path length to measure the resilience of a
city is possible but indeed not convenient, since $L$ becomes infinite
as soon as there exists at least one pair of disconnected
nodes. However, the result of a shock event on a city, like an
earthquake or a flood, is often a disconnected street network, which
always has an infinite characteristic path length independently of the
actual number of pairs of sites that remain disconnected after the
event. The network efficiency, proposed in
reference~\citep{Latora2001}, is a measure which allows to overcome
the subtleties due to infinite characteristic path lengths and can be
therefore used to quantify the average reachability of the nodes even
when the network is not connected, e.g. in the case of partially
disrupted road networks after a disaster. The efficiency $e_{ij}$ of
the communication between nodes $i$ and $j$ in a generic graph is
defined as the inverse of the length of the shortest path connecting
$i$ to $j$, i.e. $e_{ij} = 1/d_{ij}$. The efficiency is minimal and
equal to $0$ when $i$ and $j$ are disconnected, i.e. when
$d_{ij}=\infty$. In the case of spatial graphs, the efficiency of a
pair of nodes is usually normalized dividing it by the Euclidean
distance between the two nodes, so that the efficiency between $i$ and
$j$ is defined as $e_{ij} = d^{eucl}_{ij}/d_{ij}$, where
$d^{eucl}_{ij}$ is the Euclidean distance between node $i$ and node
$j$. Notice that the resulting normalized efficiency is maximal and
equal to $1$ if and only if the shortest path between $i$ and $j$ runs
exactly along the direction of the geodesic which connects them.
The \textit{global efficiency} of a spatial network is defined as the
average of the normalized pairwise efficiency over all possible pairs
of nodes~\citep{Vragovic2005}:
\begin{equation}
	E = \frac{1}{N(N-1)}\sum_{i,j\in \mathcal{N}, i\neq j} \frac{d^{eucl}_{ij}}{d_{ij}}.
	\label{eq:eff_giusta}
\end{equation}
Notice that the global efficiency is normalized in $[0,1]$, since in
general the distance $d_{ij}$ between node $i$ and node $j$, which is
measured as the total length to be traversed in order to get from $i$
to $j$ using a sequence of street segments, is larger than the
Euclidean distance between $i$ and $j$, so that each term in the
summation is $\le 1$. Consequently, it is possible to compare in a
consistent way the efficiencies of two distinct graphs $G$ and $G'$,
even if they have a different number of nodes and edges.

\subsection{Construction of Hybrid Social-Physical Networks}

The street network of a city is the main physical component of the
hybrid network description of urban systems that we propose. In order
to take into account some of the social aspects of a city, we enrich
the road network adding different kinds of nodes, which represent
citizens, buildings and facilities, and different kinds of links,
which model the relationships between social and physical elements. We
call the resulting graph a Hybrid Social-Physical Network (HSPN).
%
%
The idea of modelling
cities by constructing augmented graph models which integrate
information about their main components has recently been employed in
the field of resilience assessment (see for
instance~\citet{Cavalieri2012,Franchin2013}).

According to the nature and accuracy of the information used to
augment the road network, we can actually obtain several different
HSPN representations of a urban system. A first example is the
\textit{residential HSPN}, which includes information about population
and building locations and is constructed as follows. For each
building we add a new \textit{building node}, whose coordinates are
those of the centroid of the building footprint on the map. Then, each
building node is connected to the road network by means of a new
\textit{doorstep edge} orthogonal to the street segment closest to the
building. The other endpoint of a doorstep edge, called a
\textit{doorstep node}, is chosen to be either one of the existing
crossings in the street network or a newly ad-hoc added
node~\footnote{The choice of doorstep nodes was made by hand in order
  to guarantee, at the same time, that the direction of doorstep edges
  remains as close as possible to the direction orthogonal to the
  closest street segment and that only a small number of extra nodes
  were actually added to the existing road network.}. 

\begin{figure}
	\begin{center}
  	  \includegraphics[trim=3cm 3cm 3cm 3cm, clip=true, angle=90,width=3in]{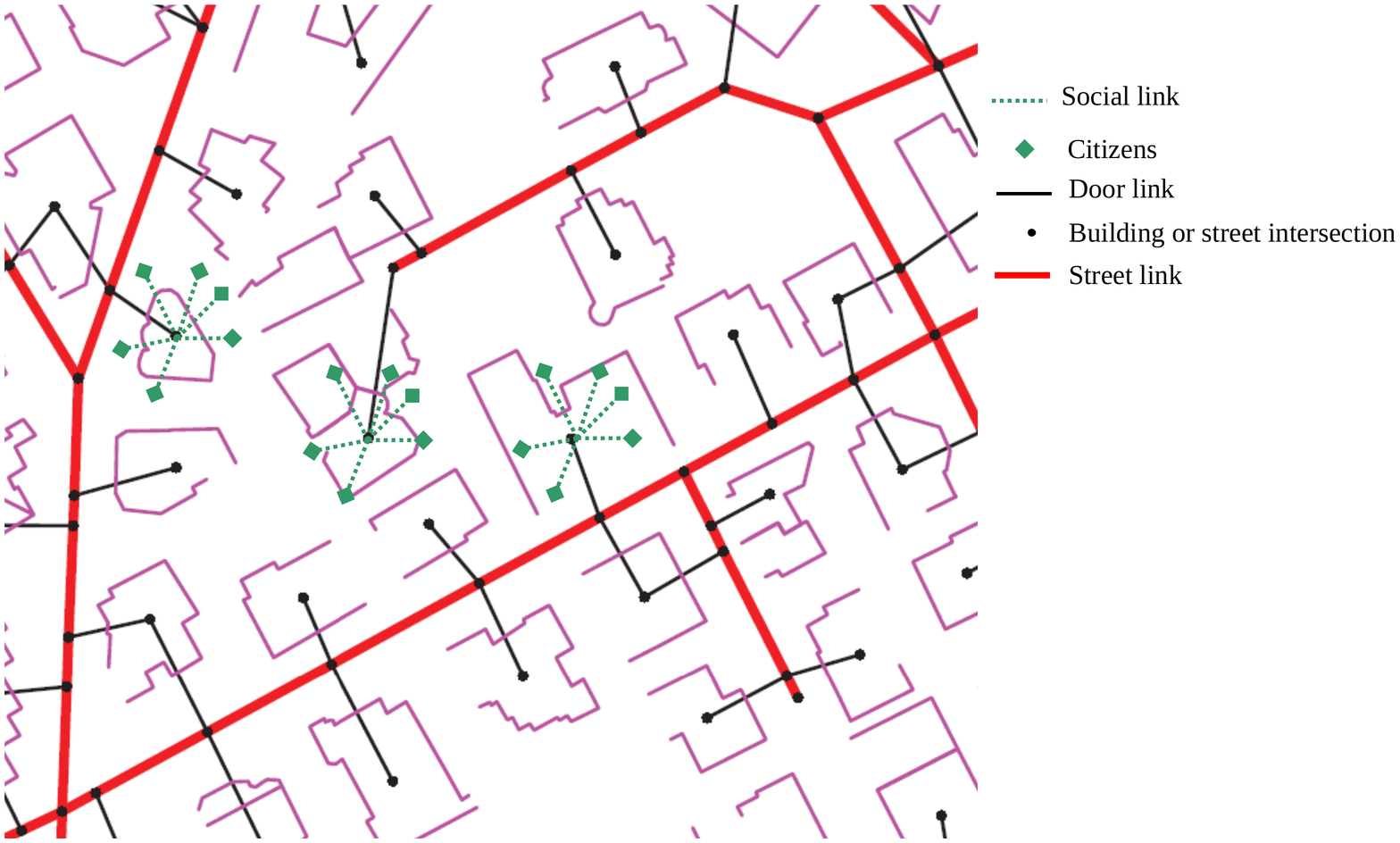}
	\end{center}
  \caption{Network representation of a city. Each building is
    associated to a new node and is connected to the road networks by
    means of a doorstep edge incident on a doorstep node. Also, each
    citizen living in a building is represented as a virtual (green)
    node attached to the building. The distance between two citizens
    living in buildings incident on the same doorstep node is set to
    zero.}
	\label{fig:nodes}
\end{figure}

The construction of the residential network is illustrated in
Figure~\ref{fig:nodes}.In this case the HSPN consists of two sets of
nodes and two set of edges, namely
\begin{itemize}
\item The set of intersection nodes, which we called $\mathcal{N}$.
\item The set of building nodes, hereafter referred as $\mathcal{B}$.
\item The set of street segments, which we called $\mathcal{V}$.
\item The set of doorstep edges, denoted by $\mathcal{V_B}$
\end{itemize}

With a little abuse of notation, we include in the set $\mathcal{N}$
the doorstep nodes, besides some of them were not initially present in
the road network and have been added to the street network just to
attach doorstep edges.  The street network augmented with the set of
building nodes and doorstep edges is denoted by
$\mathbb{G}_\mathcal{B}\mathcal{(N \cup B, V \cup V_B)}$. In order to
correctly take into account the mutual reachability of citizens, each
building is also attached to a set of \textit{citizen nodes}. These
are just virtual nodes introduced to model the relationships between
citizens and the buildings in which they live. By definition, the
distance between two citizen nodes is equal to the distance, on the
road network, between the doorstep nodes of the buildings in which
they live. Consequently, if two citizen nodes are attached to the same
building or to two separate buildings incident on the same doorstep
node then their distance is set to zero. This is clarified by
Figure~\ref{fig:nodes}.
Both the social and the physical part of the HSPN can be even more
complex, and a procedure similar to that used to construct residential
HSPNs can be employed to augment the street network with additional
information.  For instance, if we consider the street network
together with residential and commercial buildings, we can assess the
capability of citizens to reach goods supplies. The corresponding
graph is denoted by $\mathbb{G}_\mathcal{G}\mathcal{(N \cup B \cup G,
  V \cup V_B \cup V_G)}$, where $\mathcal{G}$ is the set of nodes
representing commercial buildings and $\mathcal{V_G}$ is the set of
logical links connecting this commercial buildings with the adjacents
street segments. We call this graph a \textit{goods HSPN}. In order to
quantify its relative importance, we associate to each goods building
$i$ a weight $G_i$ which is proportional to the amount of goods it
makes available to citizens.

Similarly, we can also construct a \textit{service HSPN} to assess the
capability of citizens to access public services, like schools,
hospitals and other public infrastructures. In this case the HSPN is
denoted by $\mathbb{G}_\mathcal{S}\mathcal{(N \cup B \cup S, V \cup
  V_B \cup V_S)}$, where $\mathcal{S}$ is the set of nodes
representing public buildings and $\mathcal{V_S}$ is the set of
logical links connecting public buildings to intersection nodes. As in
the case of goods HSPN, the relative importance of a service $i$ in
terms of quantity and/or quality of service provided to citizens is
encoded in a weight $S_i$.

We notice that this methodology allows to construct many other HSPNs
corresponding to different lifelines. For instance, information about
the electric grid network and the water supply/sewerage network can be
included in the model, to investigate the capability of the urban
system to provide citizens with electricity and water, respectively.

The potential of this approach lays in the fact that the global
performance of the HSPN associated to a given service can be
considered as a proxy of the accessibility of that service by the
citizens, i.e. as a measure of the quality of the service provided.
Thus, it is possible to quantify and compare the performance of the
HSPN in distinct configurations of the physical networks, even if some
infrastructures are not present in one of the configurations. By
performing this analysis before and after the reconstruction which
follows an hazardous event, which damages some of the pre-existent
physical components of the urban system, we can quantify how the
quality of public services provided to citizens has been affected,
although the physical systems are rebuilt and rearranged in a new
configuration. Thus, we believe that the HSPN approach is an effective
methodology to provide a quantitative assessment of the resilience of
a urban system ---as requested by the engineering
approach--- based on a systemic perspective that properly takes into
account the social aspects of the system and the perceived efficiency
of the city as a whole ---as indicated by the ecosystem
approach.

\subsection{Efficiency of Hybrid Social-Physical Networks}
\label{sec:performance}

We define here some efficiency metrics for HSPNs which are inspired by
the normalized global network efficiency given in
equation~\eqref{eq:eff_giusta}.  The mere definition
~\eqref{eq:eff_giusta} is inadequate to estimate the efficiency of the
city, because it does not take into account the number of people
living in each building. We can then devise a measure of efficiency of
communications between people inside buildings by summing over all
couples of inhabitants and using $d^{eucl}_{ij}/d_{ij}=1$ for couples
of inhabitants living at distance zero, i.e. assuming the maximum
efficiency in their communications.  In the same spirit of
Equation~\eqref{eq:eff_giusta}, we define the efficiency of a
residential HSPN as :
\begin{equation}
\begin{aligned}
	E_{cc} &= \frac{1}{H_{tot}(H_{tot} - 1)}\sum_{i \in \mathcal{B}}  H_i  \left(
		(H_i-1) + \sum_{j \in \mathcal{B}, j \neq i } H_j \frac{d^{eucl}_{ij}}{d_{ij}} \right)\\
	&= \frac{1}{H_{tot}(H_{tot} - 1)}\sum_{i \in \mathcal{B}}  H_i  \left(
		(h_i-1) + \sum_{j \in \left( \mathcal{B} \setminus \mathcal{I} \right) } H_j \frac{d^{eucl}_{ij}}{d_{ij}} \right),
	\label{eq:effcity_giusta}
\end{aligned}
\end{equation}
where $i,j$ are the indexes of nodes representing buildings, $H_{tot}$
is the total number of inhabitants of the city, $H_i$ is the number of
people living in building $i$, $\mathcal{B}$ is the set of nodes
representing buildings, $d_{ij}$ is the length of the shortest path
between $i$ and $j$ evaluated on graph $\mathbb{G}_\mathcal{B}$ and
$h_i$ is the number of inhabitants that live in the set $\mathcal{I}$
of buildings with zero distance to building $i$\footnote{Notice that
  the distance between two buildings is set to zero if their
  corresponding doorstep edges incide on the same doorstep node}.
This definition of efficiency for a residential HSPN, is indeed able
to quantify the mutual reachability of people living in the city. In
fact, the lower the distance among people in the augmented network,
the higher the efficiency value of the corresponding HSPN, and
vice-versa. Notice that in the summation over $j, j \neq i$ we set
$d^{eucl}_{ij}/d_{ij}=1$ for couples of buildings at zero distance.
The term $(H_i-1)$ inside the parentheses, multiplied by $H_i$, takes
into account the couples of inhabitants that live in the same building
and whose efficiency is $e_{ii} = 1$.

It is possible to define an efficiency also for goods HSPNs, by
substituting the outer summation in Equation~\eqref{eq:effcity_giusta}
with a summation over the set $\mathcal{G}$ of the buildings that
contain goods, e.g. shops and retail stores, and dividing by the
quantity $G_{tot}$ which is equal to the sum of the importance of all
goods buildings. In formula:
\begin{equation}
\begin{aligned}
	E_{cg} &= \frac{1}{G_{tot} H_{tot}}\sum_{i \in \mathcal{G}}
		 \sum_{j \in \mathcal{B} } G_i H_j \frac{d^{eucl}_{ij}}{d_{ij}}\\
 &= \frac{1}{G_{tot} H_{tot}}\sum_{i \in \mathcal{G}}  G_i  \left(
		h_i + \sum_{j \in \left( \mathcal{B} \setminus \mathcal{I} \right) } H_j \frac{d^{eucl}_{ij}}{d_{ij}} \right),
\end{aligned}
	\label{eq:effgood_giusta}
\end{equation}
where $G_i$ is an estimate of the amount of goods in the shop $i \in
\mathcal{G}$, $d_{ij}$ is the length of the shortest path between $i$
and $j$ evaluated on the graph $\mathbb{G}_\mathcal{G}$.  Notice that
Equation~\eqref{eq:effgood_giusta} measures the average of the inverse
of the normalized distance between people and goods sold in shops, and
effectively quantifies how easily citizens can access goods supplies.

Similarly, the efficiency of the service HSPN is given by the
equation:
\begin{equation}
\begin{aligned}
	E_{cs} & =
 		\frac{1}{S_{tot} H_{tot}}\sum_{i \in \mathcal{S}}
		 \sum_{j \in \mathcal{B} } S_i H_j \frac{d^{eucl}_{ij}}{d_{ij}}\\
		& =\frac{1}{S_{tot} H_{tot}}\sum_{i \in \mathcal{S}}  S_i  \left(
		h_i + \sum_{j \in \left( \mathcal{B} \setminus \mathcal{I} \right) } H_j \frac{d^{eucl}_{ij}}{d_{ij}} \right),
\end{aligned}
	\label{eq:effschool_giusta}
\end{equation}
which is just the average of the inverse of the normalized distances
between people and services. In this case, $S_i$ represents a measure
of the importance of the service $i$ (e.g. the floor area of the
building or the totol number of citizens that can potentially use the
service), $\mathcal{S}$ is the set of nodes representing service
buildings and $S_{tot}=\sum_i S_i$. In this equation the length
$d_{ij}$ of a shortest path is evaluated on the graph
$\mathbb{G}_\mathcal{S}$.

\section{Using HSPNs to quantify urban resilience}
\label{sec:hspn_resilience}

The main idea in this paper is to use the efficiency measures on HSPNs
to quantify the resilience of a urban system, by comparing their respective
values before and after a catastrophic event and during the subsequent restoration.
In this section we first report the classical definition of resilience
and introduce an ad-hoc normalization.
We then review six reconstruction
strategies which permit to simulate and trace the restoration process.

\subsection{Measures of resilience}
\label{sec:resilience_measures}

The classical approach to urban resilience is based on the definition
of a recovery function $Y(t)$, whose value at time $t$ is equal to the
measured \textit{performance} of the system at that time. If the
recovery process starts at time $t_1$ and is completed at time $t_2$,
then the resilience $R$ of the urban system is defined as the area
under the recovery curve~\citep{Reed2009}:\footnote{Some authors refer
  to the resilience as the area over the recovery
  function~\citep{Bruneau2003}}
\begin{equation}
	R=\frac{\int^{t_{2}}_{t_{1}} Y(t) dt}{t_2-t_1}.
	\label{eq:resilience_true}
\end{equation}

Notice that $R$ depends both on the time required by the recovery
process and on the reconstruction strategy adopted. By using the HSPN
model of a city, it is possible to define the performance of the urban
system at time $t$ as the ratio between the efficiency $E(t)$ of the
associated HSPN at time $t$ and the efficiency $E_0$ of the HSPN just
before the disaster occurred, namely
\begin{equation}
  Y(t) = E(t)/E_0
\end{equation}
This definition of performance requires to know the structure of the
HSPN at each time during reconstruction. However, this information
depends on several factors including the available budget and the
promptness of the reconstruction, and is usually not easy to
obtain. Consequently, we remove any explicit
dependence of the resilience on time, and use a recovery function
$Y(C)=E(C)/E_0$ defined as the ratio between the efficiency $E(C)$ of
the urban system when $C$ citizens have been relocated and the
efficiency $E_0$ immediately after the disaster.
In particular, we
make use of the normalised performance:
\begin{equation}
y(C) = \frac{Y(C)-Y(0)}{1-Y(0)}.
\label{eq:normal_perform}
\end{equation}

Notice that $y(C)=0$ in the aftermath of the disaster, i.e. when
$y(C)=1$ when $Y(C)=1$. In order to quantify the resilience of a urban
system we define the measure:
\begin{equation}
	\mathcal{R}=\frac{\int^{C_{max}}_{0} y(C) dC}{C_{max}},
\label{eq:resilience}
\end{equation}
where $C_{max}$ is the total number of people to be relocated after a
certain event. In the following we employ
Equation~\eqref{eq:resilience} instead of the classical definition of
resilience given in Equation~\eqref{eq:resilience_true},
as it permits to appreciate the recovery of a city closely after the disaster and to compare
different strategies of reconstruction after events of different magnitude.
To date, a discussion about a proper way to compute a normalized resilience is ongoing~\citep{FranchinICOSSAR}.

\subsection{Reconstruction strategies}
\label{sec:reconstruction}

We review here the six reconstruction strategies considered in the
case study reported in Section~\ref{sec:case_study}.

The first strategy, referred to as \textit{status quo down-up}, aims
at re-obtaining exactly the same configuration of buildings and
services that the city had before the catastrophic event occurred.
The reconstruction process is discretized into $n$ steps. In each step
a fraction $1/n$ of the displaced citizens (those living in the
damaged buildings) is reallocated, assuming that the buildings are
restored starting from the smallest ones and proceeding towards the
largest ones, i.e. from the cheapest to the most expensive.  Blocked
roads are recovered when the buildings that caused their interruption
are made safe or reconstructed.  In the second strategy, referred to
as \textit{status quo up-down}, the city eventually returns to the
undamaged configuration, as in the \textit{status quo down-up}, but
the restoration process starts with the largest buildings and
proceeds towards the smallest ones.

The third strategy, hereafter referred to as \textit{new sites
  down-up}, consists in reallocating part of the displaced citizens in
new residential sites.  Also this process is discretized into steps.
In the first step some new buildings are constructed in $4$ empty
areas (highlighted in figure~\ref{fig:empty}), to reallocate $20 \%$
of the displaced citizens. Then the existing buildings are restored as
in the first strategy, from the smallest to the largest ones, until
all citizens have been reallocated. Notice that in the
\textit{new sites down-up} strategy some of the existing buildings are
not ever recovered, since part of the population is reallocated in the
newly constructed buildings. To re-establish the original urban street
pattern, it has been assumed that the interrupted roads, that would
have not been recovered (since interrupted by buildings that are not
recovered), are re-established during the last step.
The fourth strategy, called \textit{new sites up-down}, is similar to
the \textit{new sites down-up} one, but, after reallocating the first
$20\%$ of citizens in the newly constructed buildings, the restoration
of damaged buildings proceeds from the largest to the smallest ones.
The \textit{new sites} strategies are adopted in
real cases to recover basic urban functionalities as fast as possible,
and have been recently employed in the aftermath of earthquakes,
e.g. in the case of L'Aquila 2009 earthquake, in Italy~\citep{CASE}.

The fifth strategy, called \textit{status quo inwards}, consists in
rebuilding the city as in its undamaged configuration, moving from the
suburbs to the centre. Finally, the \textit{status quo outwards} is
similar to \textit{status quo inwards} but the reconstruction starts
from the centre of the city and proceeds toward the suburbs.

\section{A case study: the city of Acerra}
\label{sec:case_study}

As a case study to test our methodology for resilience quantification
we considered a urban area ---the city of Acerra, in Italy--- and we
simulated several earthquake scenarios, evaluating the damage caused
by each earthquake and the ability of different reconstruction
strategies in restoring the pristine performance of the urban system,
measured in terms of the efficiency of the corresponding residential,
goods and services HSPN.

\begin{figure}
	\begin{center}
  	  \includegraphics[angle=0,width=3in]{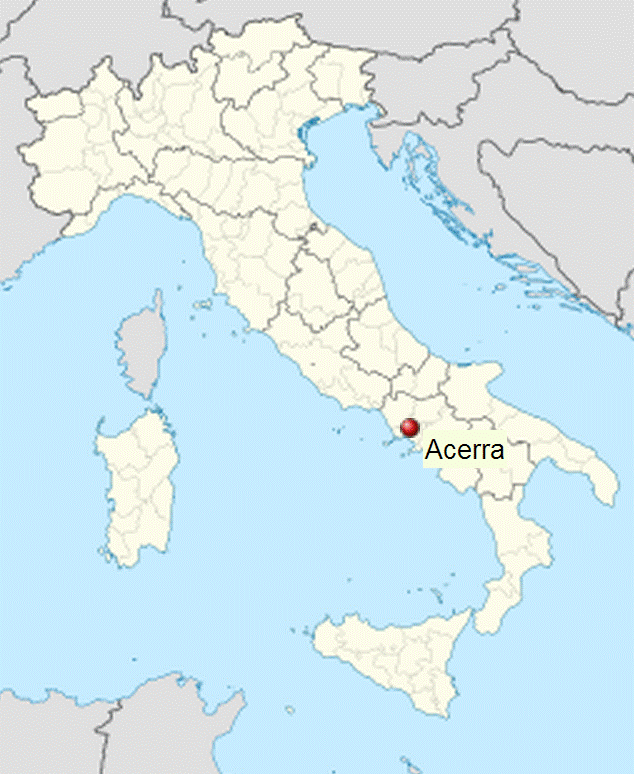}
	\end{center}
	\caption{Acerra location.}
	\label{fig:Acerralocation}
\end{figure}

\begin{figure}
	\begin{center}
  \includegraphics[trim=3cm 9cm 2.5cm 7.5cm, clip=true, angle=0,width=3in]{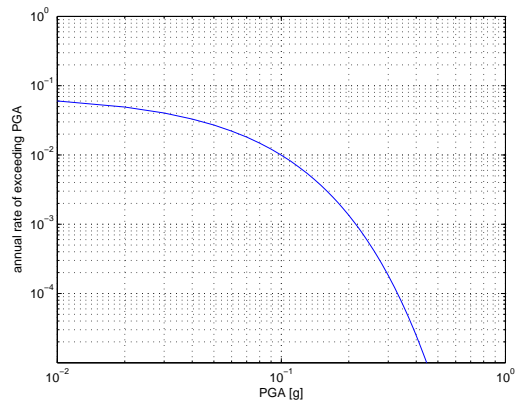}
	\end{center}
	\caption{Seismic hazard: annual rate of eartquakes exceeding a
    certain value of PGA as a function of PGA.}
	\label{fig:hazard_acerra}
\end{figure}

Acerra is a medium-sized city in the Province of Naples, in Italy,
about $20\mathrm{km}$ north-east of Naples (Figure
\ref{fig:Acerralocation}); its foundation dates back to as early as
400 BC, which makes it one of the oldest cities in that region. The
urban configuration, typical of many other medium-sized cities in
Italy, is characterised by a dense historical centre, where most of
the buildings are ancient masonry buildings, surrounded by more recent
urban expansion areas, mainly consisting of reinforced concrete
buildings; the built-up area is surrounded by a countryside area,
where a large industrial settlement is also located.  The whole
territory extends over a surface of about $54 \mathrm{km}^2$ and its
population is estimated in about $55.000$ inhabitants.
Acerra is prone to seismic risk, due to its closeness to the
seismogenic areas of the Appennines, which are just about one hundred
kilometres away from the city center.  Figure~\ref{fig:hazard_acerra}
reports the seismic hazard in terms of annual rate of occurrence of
events with a Peak Ground Acceleration (PGA) larger than a certain
value. Furthermore, the city is exposed to flood risk, due to the Regi
Lagni river which borders the built-up area, and to industrial risk,
due to the infrastructures and factories located in the industrial
area around Acerra. In the following we will focus on seismic risk
alone.

\subsection{Numerical simulations}

In order to study the resilience of Acerra, we extracted detailed
information about the street network organisation, location of
buildings and building typology. Then, we
constructed three different HSPNs (namely, a residential HSPN, a goods
HSPN and a service HSPN where the set of services was restricted to
schools), and we simulated earthquake scenarios corresponding to
increasing PGA values. By integrating this information with an ad-hoc
fragility model, we simulated building failures due to each earthquake
and then we analysed the subsequent reconstruction process, by
comparing the performance of six different reconstruction strategies.
We provide here the details of the simulations, and we discuss
afterwards the results.

\begin{figure*}
	\begin{center}
  	  \includegraphics[angle=0,width=0.8\textwidth]{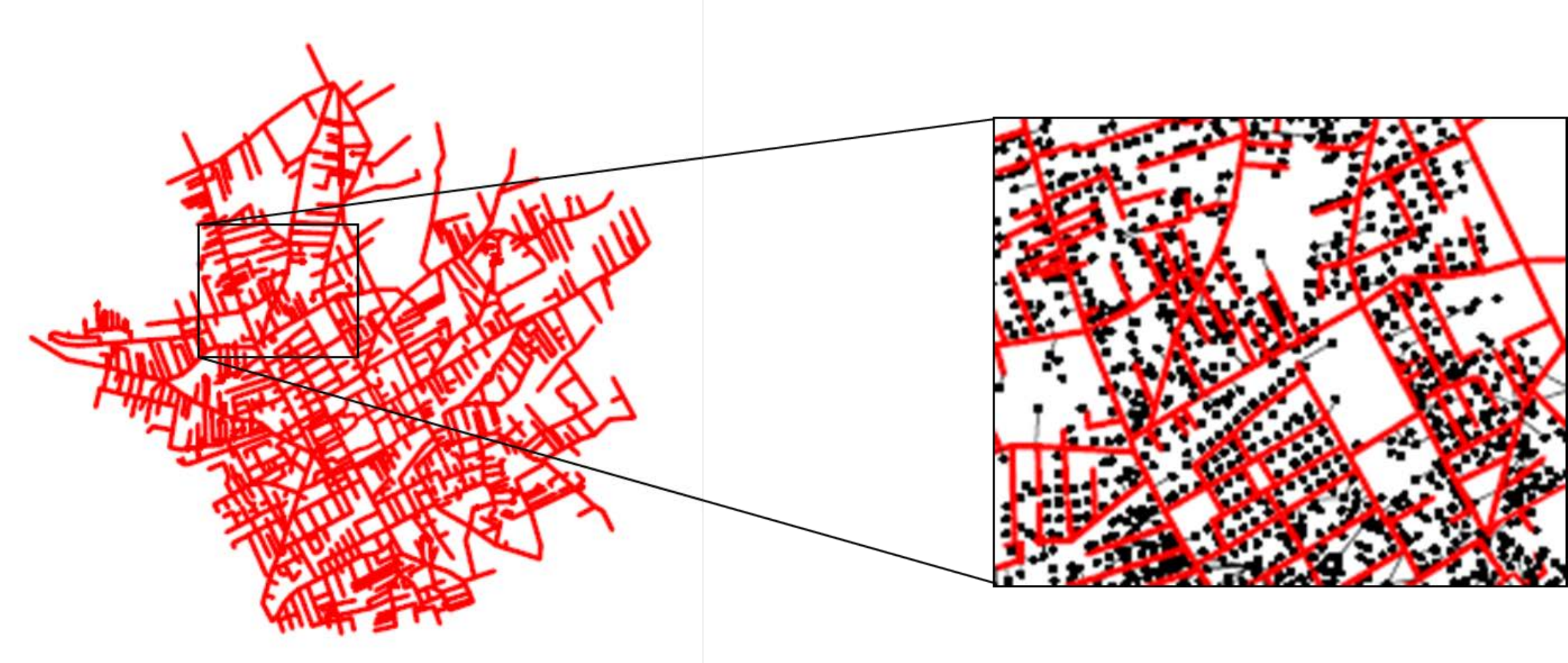}
	\end{center}
\caption{Buildings (black points) and street patterns (bold lines) network for the city of Acerra.}
	\label{fig:Acerrarete}
\end{figure*}

\medskip
\noindent
\textbf{Data acquisition.} We used GIS software to collect and integrate information about
the street network and the location and typology of buildings in
Acerra. We divided buildings into structural typologies and we
measured the total floor area of each typology, which was lately used
to estimate the number of citizens living in each residential building
and the relative importance of retail shops and services. In
Figure~\ref{fig:Acerrarete} we show the street network of Acerra, in
which the positions of buildings are reported as black squares.

\medskip
\noindent
\textbf{Fragility model.} We employed an ad-hoc model to estimate the
probability that each building will be damaged by an earthquake of a
certain intensity and the probability that a damaged building would
also interrupt the transit along the streets to which it is adjacent.
Concerning building damage, we modelled the probability that the
building would exceed the ``onset of damage'' (and thus would be
considered unfit for occupation or use) due to an earthquake of PGA
equal to $x$ through a log-normal distribution function:
\begin{equation}
	P_b\left( x; \mu, \sigma \right) = \Phi \left( -\frac{\ln
    x-\mu}{\sigma} \right).
\label{eq:fragility}
\end{equation}
The parameters $(\mu,\sigma)$ have been set equal to $(-1.03,0.35)$
for masonry buildings and equal to $(-0.91,0.29)$ for reinforced
concrete buildings, according to reference~\citep{ahmad2011}. Notice
that the adopted fragility model does not consider building height as
a parameter and does not take into account failure correlation.

If a building is over the onset of damage, then its failure could also
make inaccessible the streets adjacent to it, either because of
building debris fallen on the road or because of access restrictions
imposed for safety reasons. This road interruption probability is
defined as:
\begin{equation}
	P_r \left(h,l\right)=
	\begin{cases}
		1		& \text{if} \quad h \ge l,\\
		\frac{h}{l} 	& \text{otherwise},
	\end{cases},
	\label{eq:functional}
\end{equation}
where $h$ is the height of the building and $l$ is the width of the
road. Notice that the higher the building and the narrower an adjacent
road, the higher the probability for that road to be made inaccessible
if the building is over the onset of damage.  If a street segment is
inaccessible, it is removed from the street network. In general, the
removal of street segments has negative effects on the overall
reachability of the street network (and of the HSPN obtained from the
same network), and could also cause the fragmentation of the street
network into several isolated components, separated from each other.

\medskip
\noindent
\textbf{HSPN parameters.} The definitions of efficiency for
rsidential, goods and service HSPNs given in
Equations~\eqref{eq:effcity_giusta},~\eqref{eq:effgood_giusta}
and~\eqref{eq:effschool_giusta}, depend on the quantification of the
number of citizens living in each residential building ($H_i$ and
$h_i$) and of the relative importance of stores ($G_i$) and service
buildings ($S_i$). The number of inhabitants of residential buildings
was estimated by considering the average density of inhabitants per
square meter, obtained by dividing the total number of citizens in
Acerra (i.e., 55.000) by the total number of square meters in
residential buildings. This yields a value of one inhabitant per 30
square meters.  Concerning the goods HSPN, we considered the total
area of each store as a proxy of its importance. Finally, we
considered services HSPN restricted to schools and students, and we
considered the total area of a school as a proxy for its relative
importance $S_i$.

\medskip
\noindent
\textbf{Earthquake simulation.} We employed Monte Carlo techniques to
simulate several earthquakes scenarios corresponding to increasing PGA
values. In each scenario, we computed the probability for each
building of being beyond the onset of damage (according to
Equation~\eqref{eq:fragility}), and the corresponding probability for
roads to be made inaccessible by damaged buildings (according to
Equation~\eqref{eq:functional}). This resulted, for each realisation,
in a certain number of citizens to be relocated because of the damage
inflicted by the earthquake to residential buildings, and to a set of
street segments to be removed from the road network and from the HSPNs
since they were made unusable by damaged buildings, respectively. For
each simulated scenario we constructed the corresponding residential,
goods and schools HSPNs, and we evaluated the damage inflicted to the
urban system as the difference between the efficiency of the HSPN
before the simulated earthquake and the efficiency of HSPNs right
after the earthquake occurred.

\medskip
\noindent
\textbf{Recovery simulation.} For each earthquake scenario, we
simulated the six different reconstruction strategies detailed in
Section~\ref{sec:reconstruction}. During reconstruction, buildings
were progressively put back in place, citizens were relocated and
damaged street segments restored. We used
Equation~\eqref{eq:normal_perform} as a recovery function and
Equation~\eqref{eq:resilience} to compare the
recovery of the urban system due to the implementation of different
strategies.

\subsection{Results}

\begin{figure}
	\begin{center}
	\includegraphics[trim=0cm 7cm 0cm 7cm, clip=true, angle=0,width=3in]{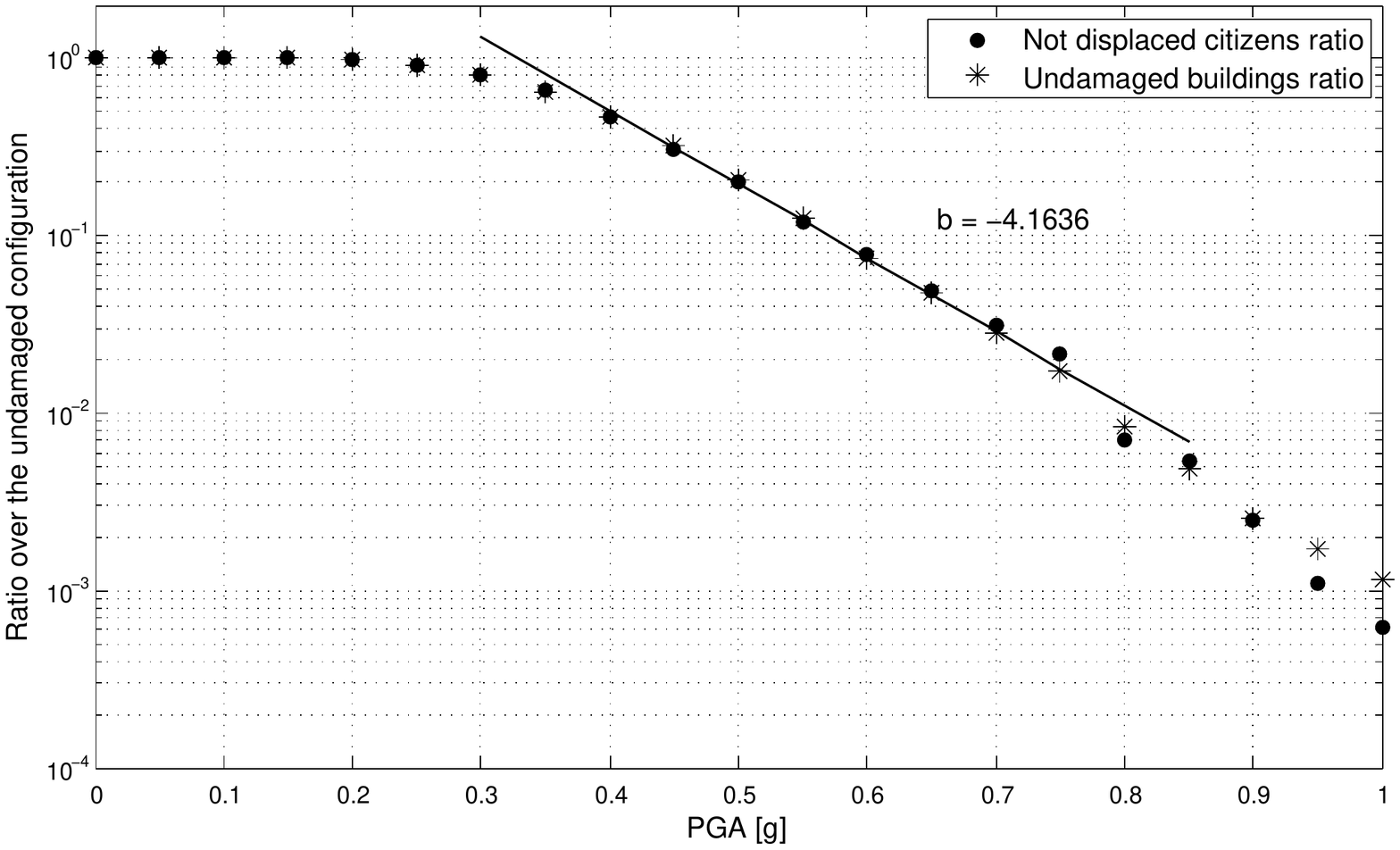}
	\caption{The number of undamaged buildings and the total number of
    not displaced citizens as a function of PGA. Notice that both
    quantities decrease exponentiallyu with PGA, with an exponent
    $b\simeq 4$. }
	\label{fig:Build-cit_PGA}
	\end{center}
\end{figure}

\begin{figure}
	\begin{center}
	\includegraphics[trim=0cm 7cm 0cm 7cm, clip=true, angle=0,width=3in]{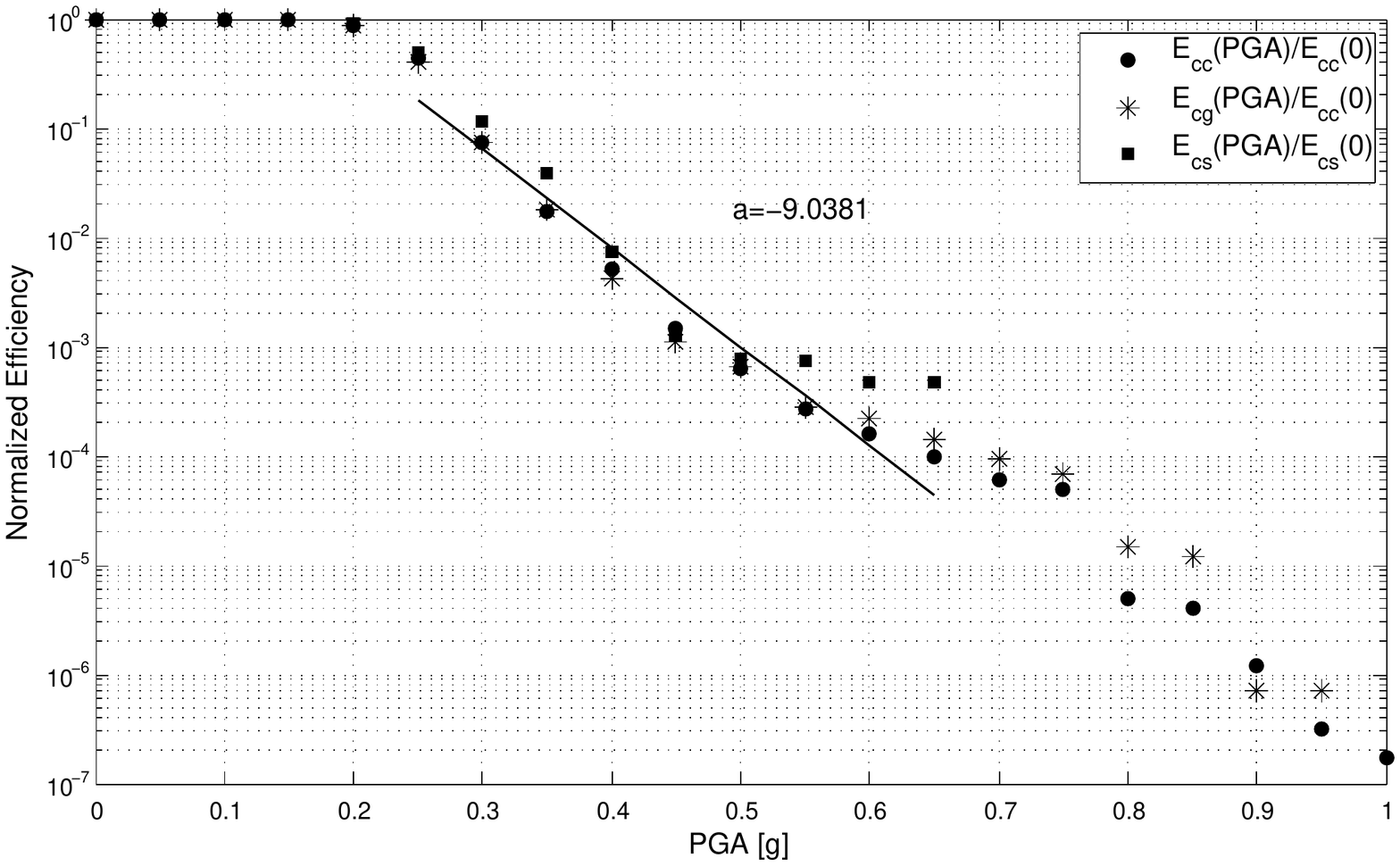}
	\caption{The values $E_{cc}$, $E_{cg}$ and $E_{cs}$ divided by the
    original efficiency in the pre-shock networks as a function of
    PGA.}
	\label{fig:E_PGA}
	\end{center}
\end{figure}

\begin{figure}
	\begin{center}
  \includegraphics[trim=3.5cm 9.5cm 3.4cm 9.5cm, clip=true, angle=0,width=3in]{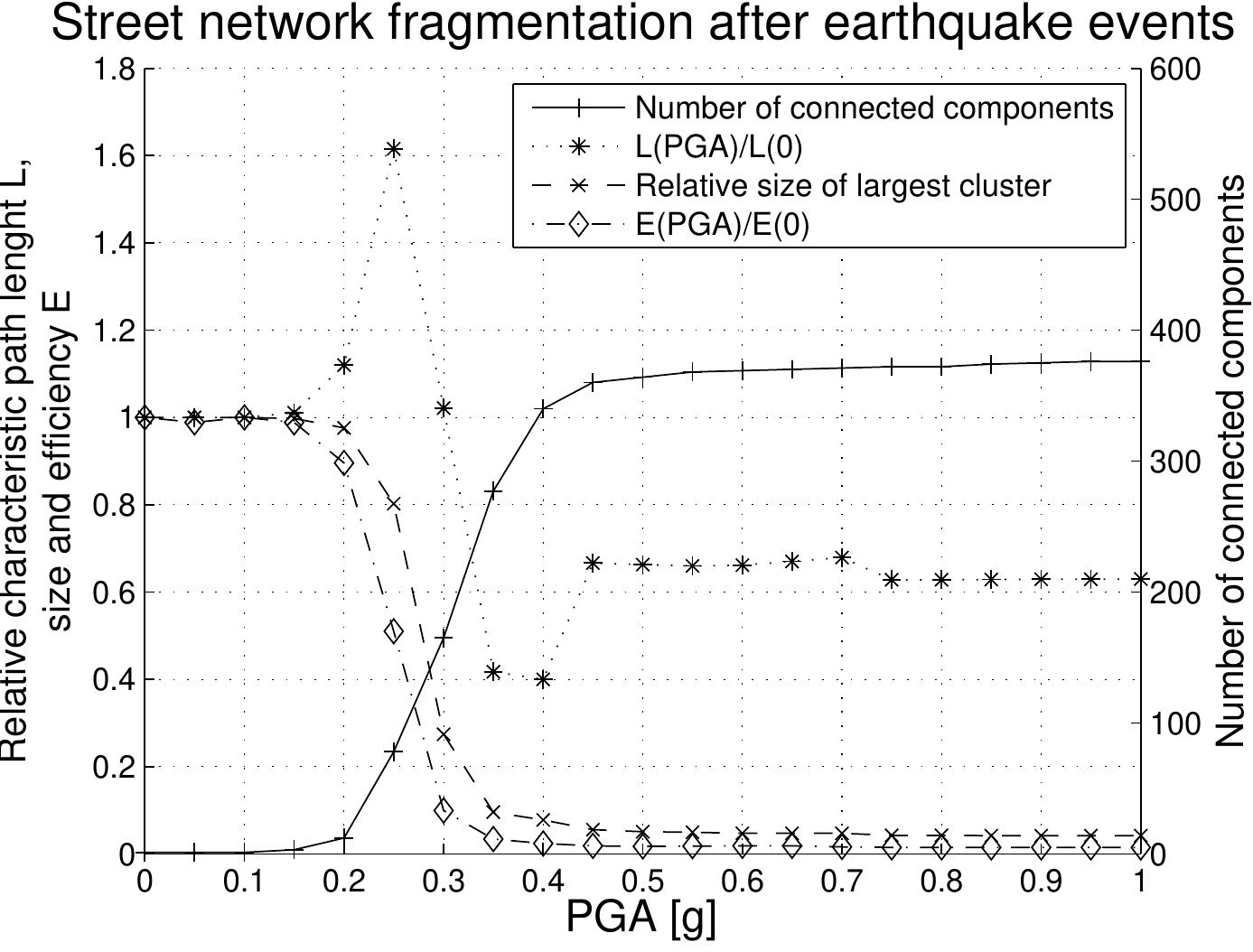}
	\caption{Street network fragmentation under earthquake event. The
    number of connected components (solid line), the normalized
    characteristic path length of the largest component $L(PGA)/L(0)$
    (dotted line), the normalized size of the largest component
    $S(PGA)/S(0)$ (dashed line), and the normalized efficiency
    eq. \eqref{eq:eff_giusta} of the graph
    $\mathbb{G}(\mathcal{N},\mathcal{V})$ (dash-dotted line) are
    plotted as function of increasing intensity of earthquake,
    measured in unit of PGA.}
	\label{fig:Street_PGA}
	\end{center}
\end{figure}

We considered several earthquake scenarios, corresponding to PGA
values in the range $[0.05g, 1.0g]$, and we evaluated three sets of
measures for each scenario, comparing their values with the undamaged
configuration.  In particular, we computed:
\begin{enumerate}
\item The number of undamaged buildings and the number of not
  displaced people as a function of PGA, whose values are reported in
  figure \ref{fig:Build-cit_PGA} .

\item The values of $E_{cc}$, $E_{cg}$ and $E_{cs}$ corresponding,
  respectively, to residential, goods and schools HSPN. The results
  are reported in Figure~\ref{fig:E_PGA}.
  
\item Several metrics to quantify the performance of the damaged
  street network, including the number of connected components, the
  number $S$ of nodes in the largest component and the characteristic
  path length $L$ of the largest connected component. The results are
  reported in Fig.~\ref{fig:Street_PGA}.
\end{enumerate}

Notice that all the results shown in
Fig.~\ref{fig:Build-cit_PGA},~\ref{fig:E_PGA} and~\ref{fig:Street_PGA}
are normalised with respect to the initial (undamaged) configuration.
The absolute values corresponding to the initial configuration are
reported in Table~\ref{tab:absval}.

\begin{table}[tp]%
\caption {Characteristics of the undamaged configuration.}
\label{tab:absval}\centering %
\begin{tabular}{lr}
\\
$E_{cc}$ & $0.746$ \\
$E_{cg}$ & $0.252$ \\
$E_{cs}$ & $0.313$ \\
\\
undamaged buildings & $3493$ \\
not displaced citizens  & $55000$ \\
\\
Number of connected components  & $1$ \\
Characteristic path length (m) & $L(0) = 1798$ \\
Nodes belonging to the largest component & $S(0) = 4638$\\
\end{tabular}
\end{table}

Figure \ref{fig:Build-cit_PGA} suggests that the number of undamaged
buildings and the number of not displaced people have the same
exponential decay with PGA, and that the corresponding exponent value
is $b\simeq-4$. This high correlation between undamaged buildings and
not displaced people is mostly due to the fact that we assumed a
constant value for the number of inhabitants per square meter. Also,
we could have expected a different decay exponent if we had considered
that people could have also been displaced from undamaged buildings,
e.g. because of unavailability of basic support
services~\citep{Cavalieri2012}.  We observe a faster exponential
decay, with exponent $b\simeq-9$, for $E_{cc}$, $E_{cg}$ and $E_{cs}$
for values of PGA between $0.25g$ and $0.6g$, as shown in
Figure~\ref{fig:E_PGA}.  These results confirm that the efficiency of
HSPNs can be reliably used as a proxy of the quality of a urban
system.

Furthermore, from Figure~\ref{fig:Street_PGA} we note that the
behaviour of the street network is consistent with a percolation
transition, indicating the existence of a critical PGA range around
$0.25g$ beyond which the street network is broken into many parts and
does not exhibit a giant connected component any
more~\citep{stauffer,Dorogovtsev2001}. This is due to the fragility
functions adopted to simulate building damage, which have a
significant increase in the collapse probability for values of PGA in
the range $[0.20g, 0.30g]$. 
It is also evident from Figure~\ref{fig:Street_PGA} that the relative
characteristic path length has a peak corresponding to PGA=$0.25g$. We
believe that this abrupt increase is due to the failure of most
short-cut roads.  Although we cannot claim that the critical PGA value
for the percolation transition is exactly at PGA$=0.25$ (a correct
estimation of this threshold would require more fine-grained
calculations), we notice that such a value of PGA would most probably
damage the majority of the masonry buildings, which are placed in the
center of Acerra, and would consequently fragment the city into a
large number of small connected components. Therefore, it is actually
the high density of masonry buildings located in the city center which
causes the sudden fragmentation of the street network.

\begin{figure*}
	\begin{center}
	\subfloat[][]{\includegraphics[angle=0,width=0.40\textwidth]{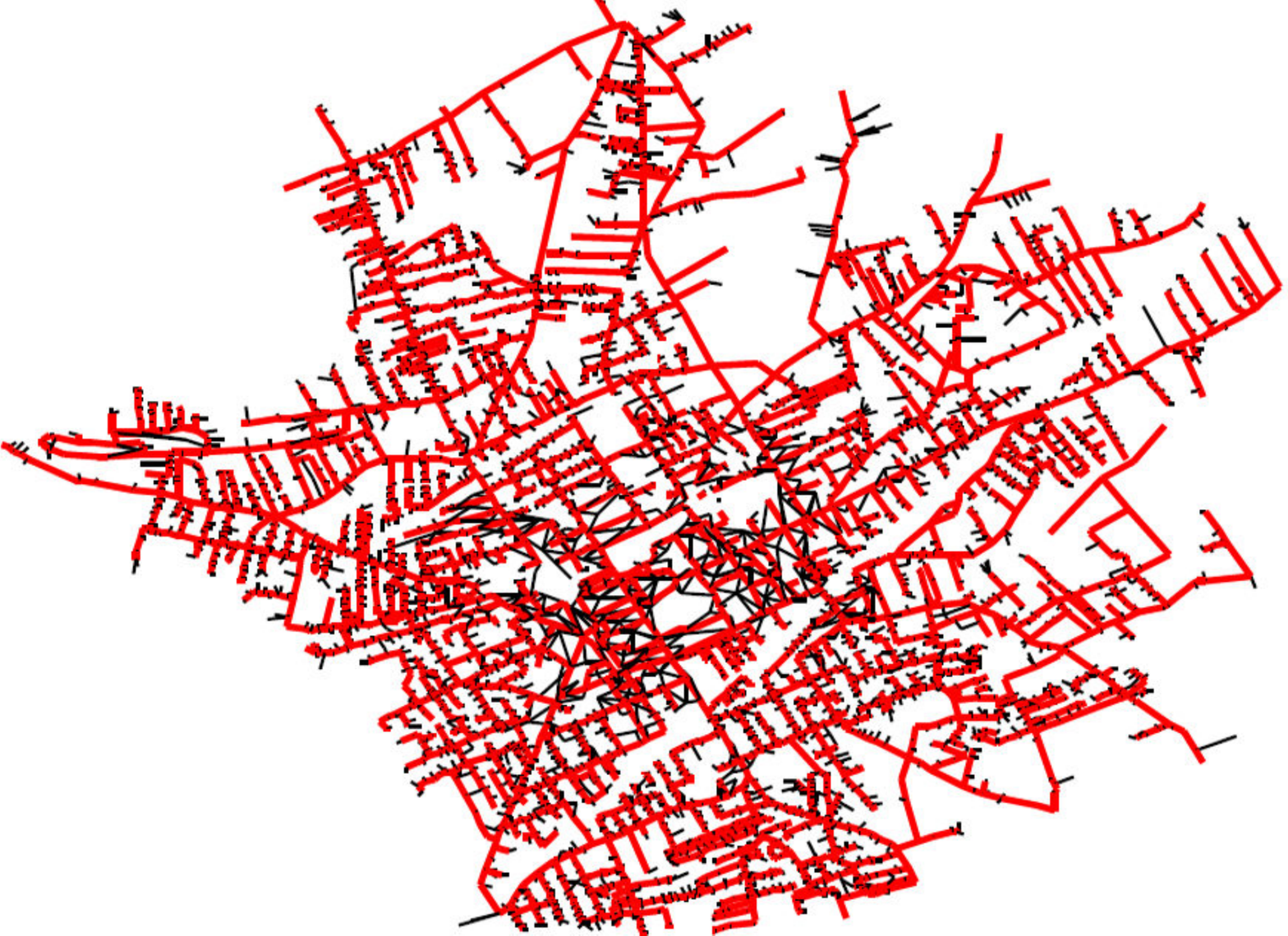}} \hfill
	\subfloat[][]{\includegraphics[angle=0,width=0.45\textwidth]{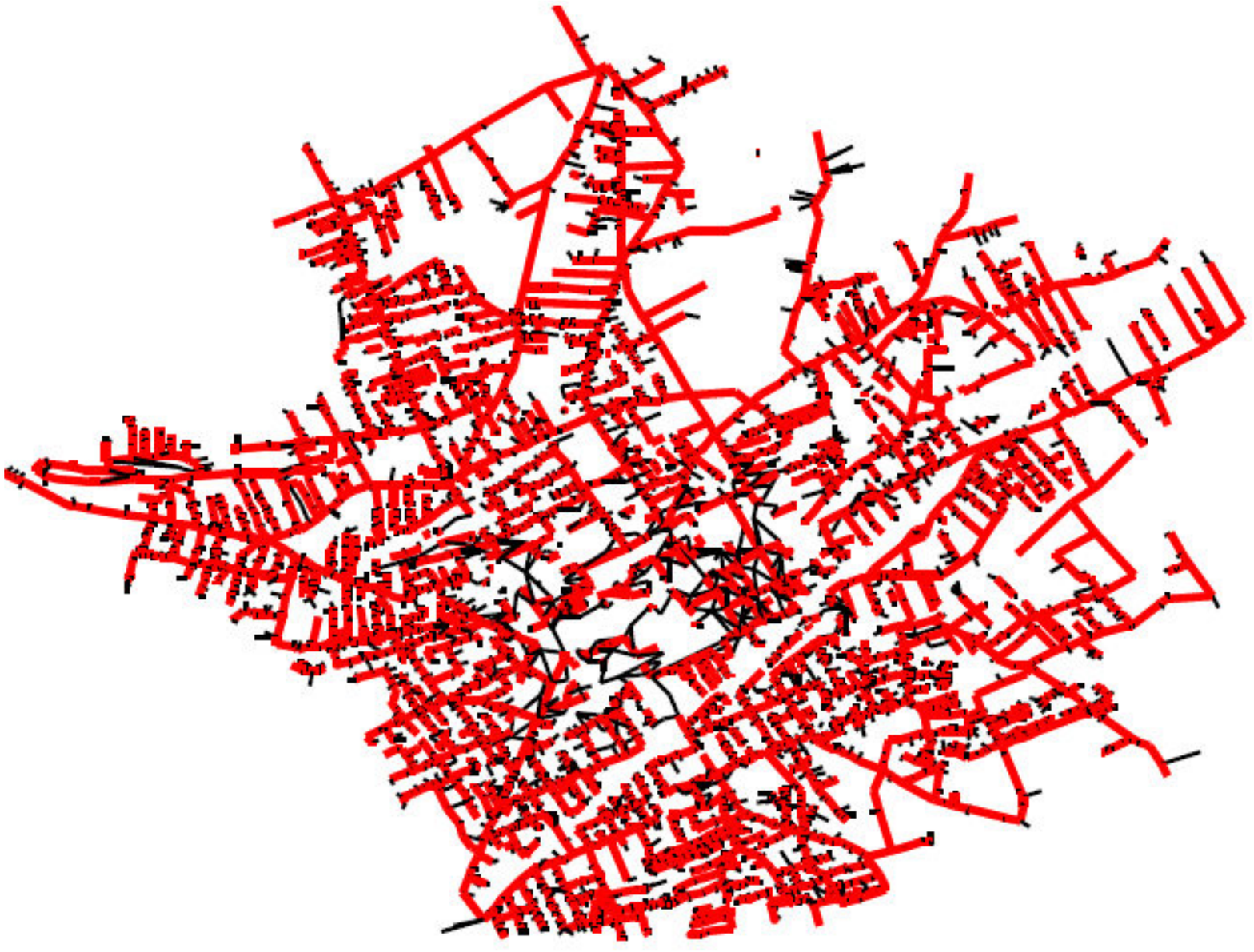}} \\
	\subfloat[][]{\includegraphics[angle=0,width=0.45\textwidth]{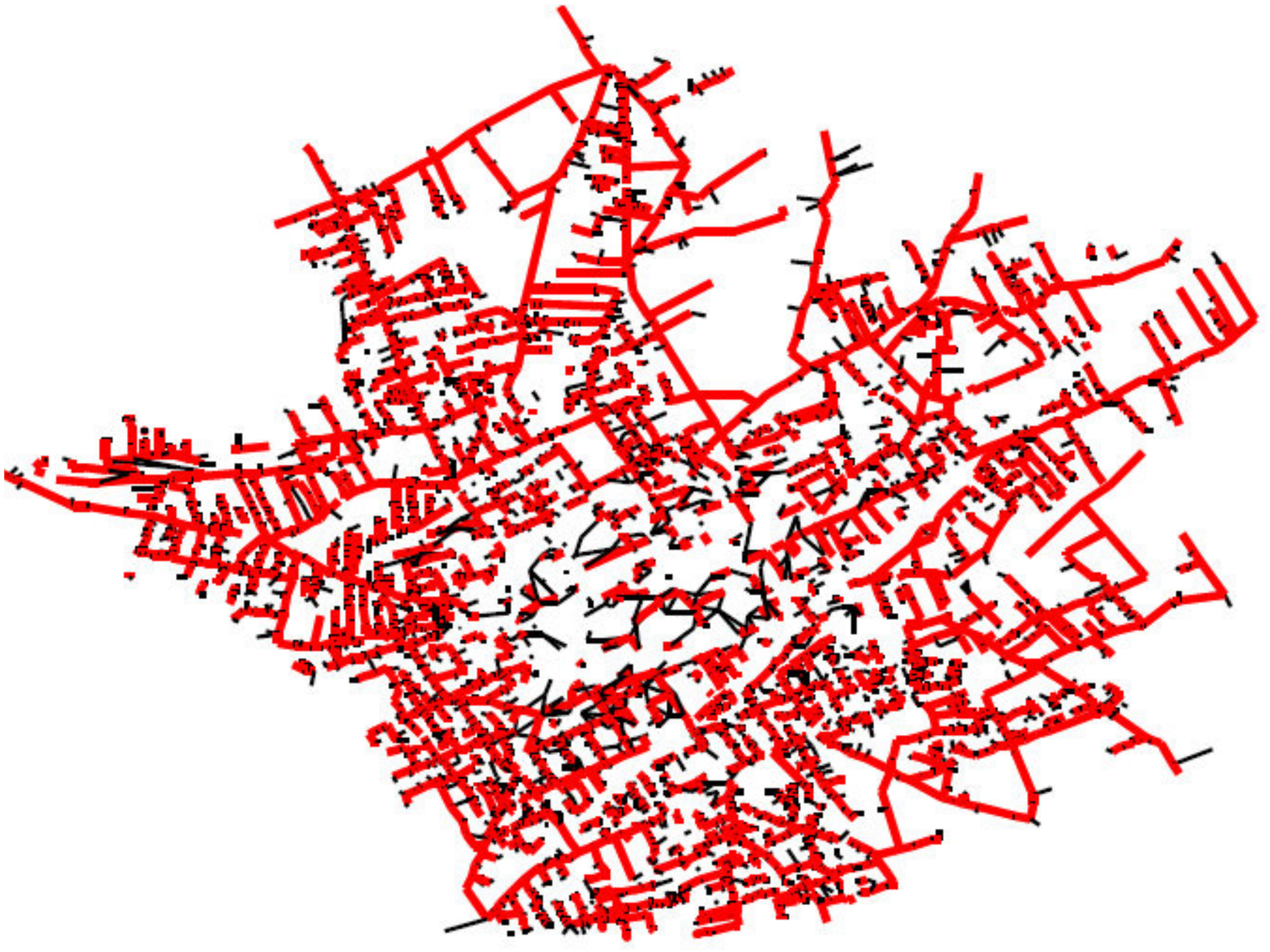}} \\
	\caption{Configuration of the simulated damaged physical network
    after simulated earthquake scenarios of PGA $0.2 g$ (a), $0.2
    g$(b) and $0.3 g$(c).}
  \label{fig:dam_scenarios}
	\end{center}
\end{figure*}

\begin{figure}
\begin{center}
\subfloat[][]{\includegraphics[trim=0cm 5cm 0cm 8cm, clip=true, angle=0,width=3in]{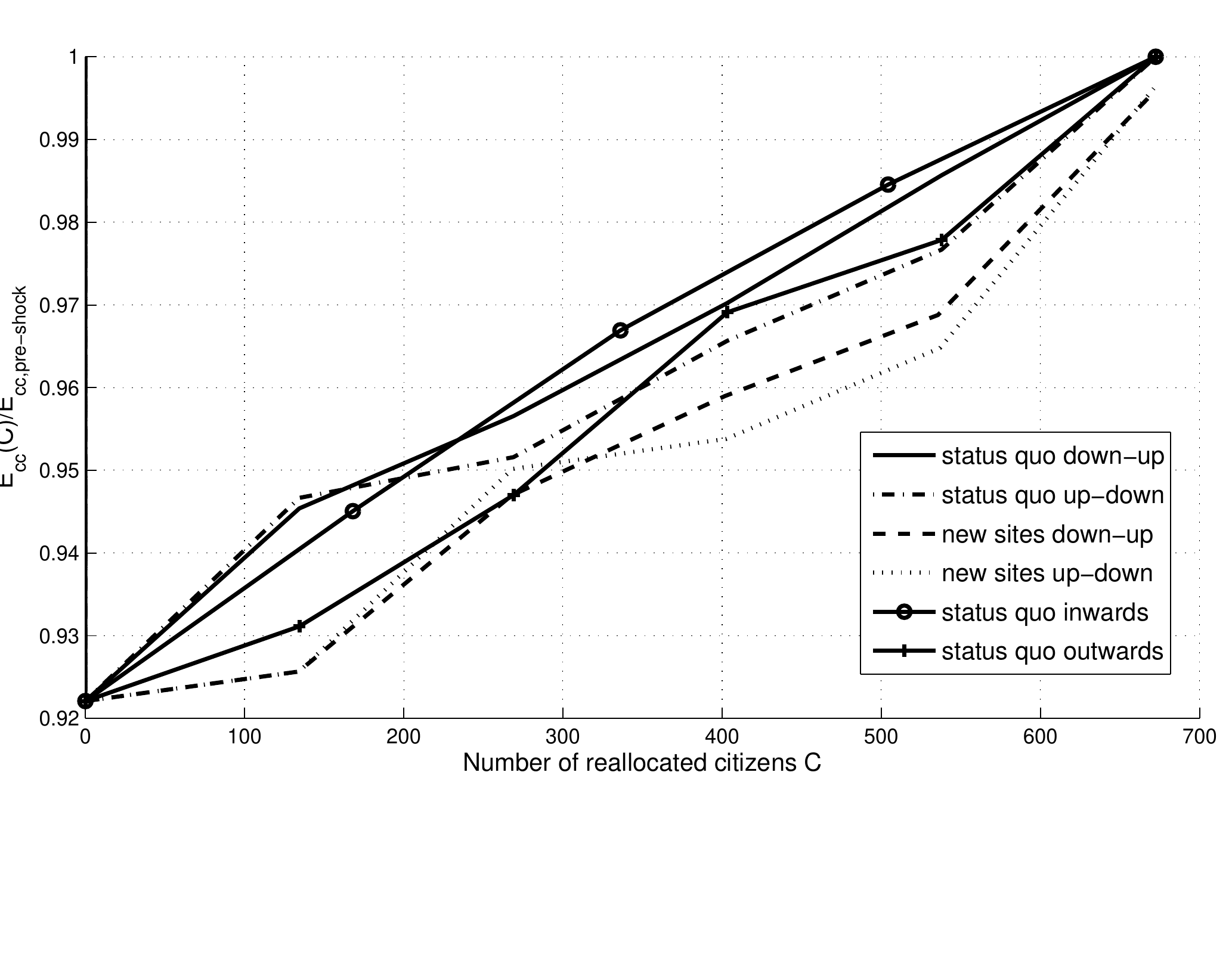}} \hfill
\subfloat[][]{\includegraphics[trim=0cm 5cm 0cm 8cm, clip=true, angle=0,width=3in]{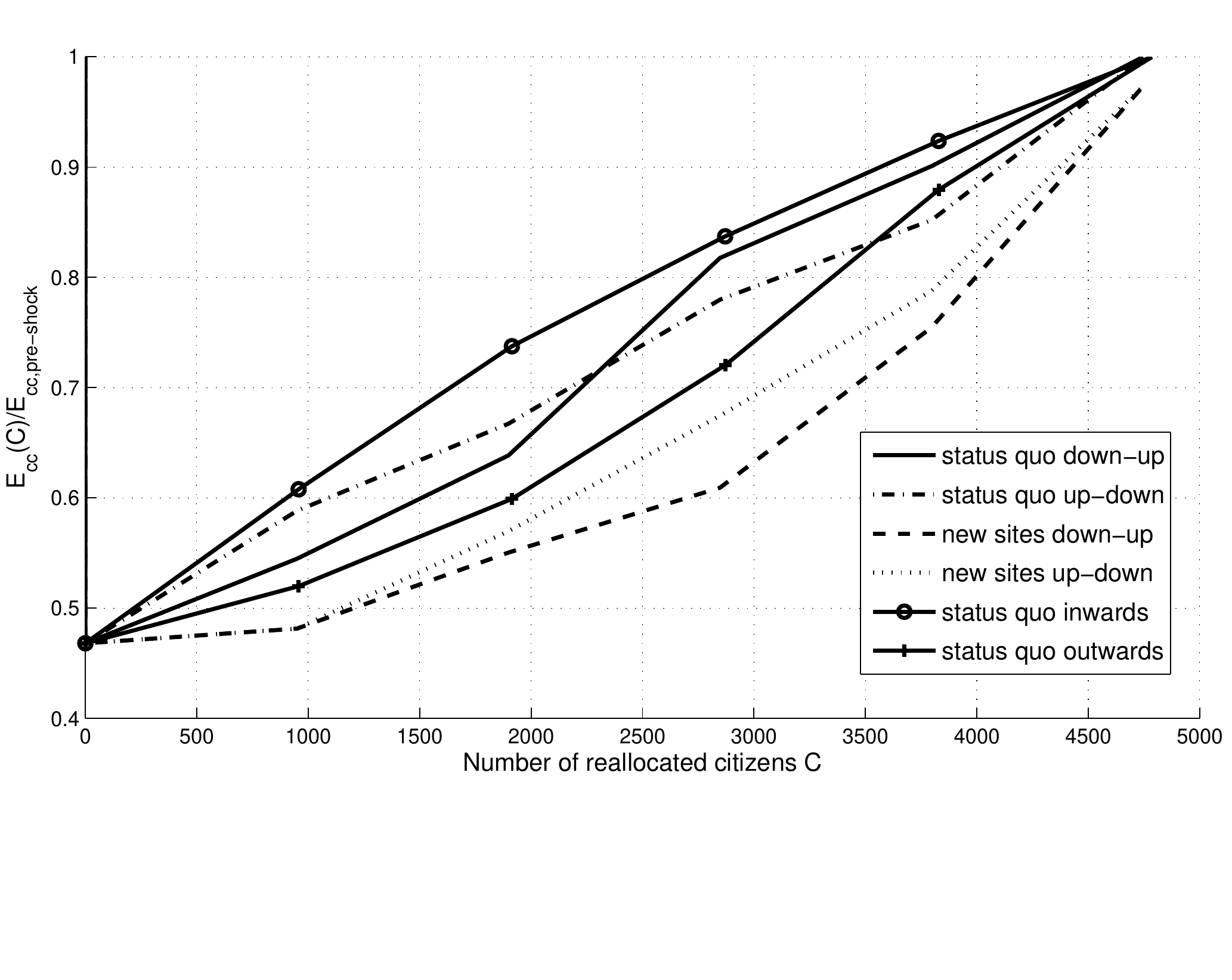}} \\
\subfloat[][]{\includegraphics[trim=0cm 5cm 0cm 8cm, clip=true, angle=0,width=3in]{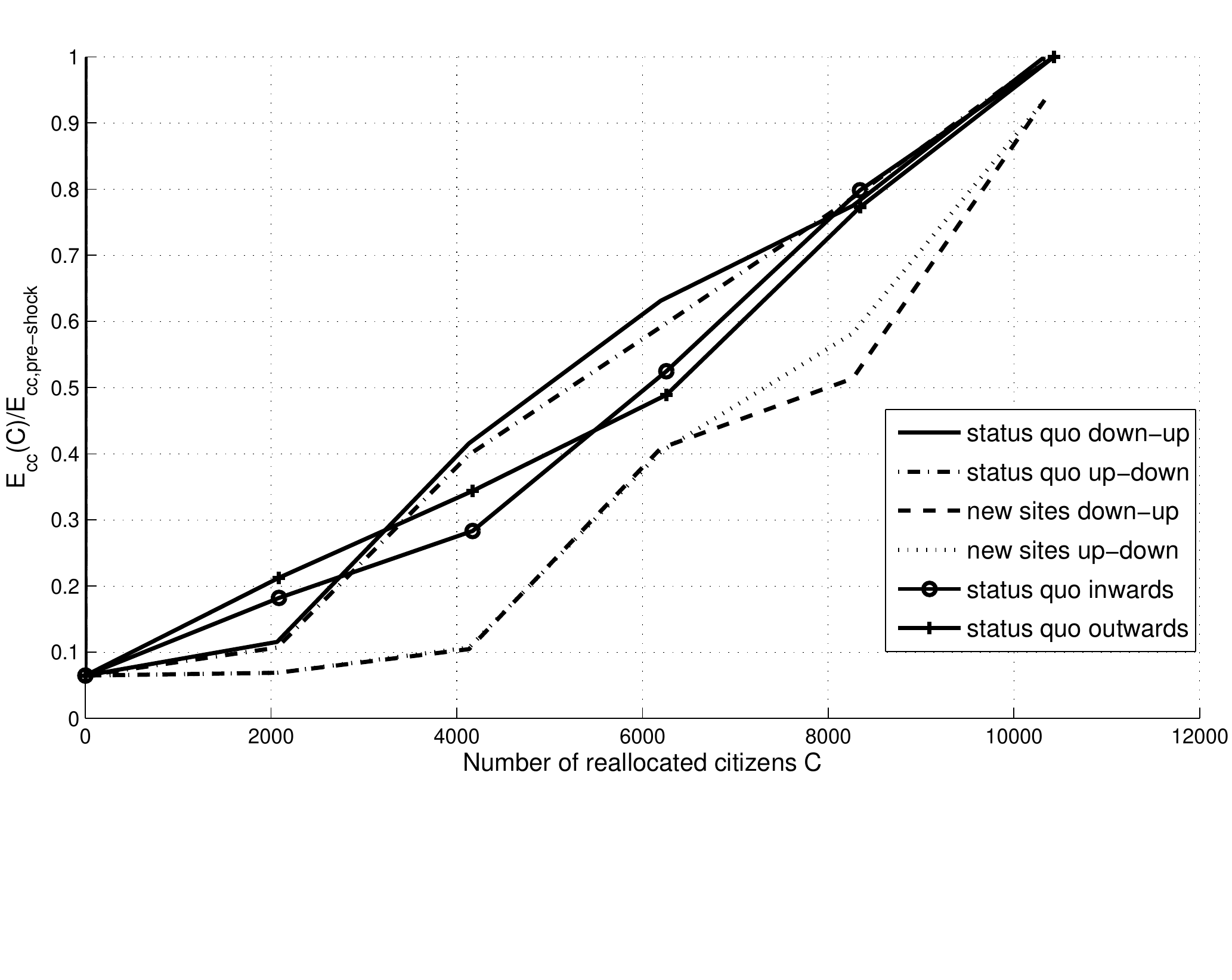}} \\
\caption{The value of the efficiency $E_{cc}(C)$ of the residential
  HSPN as a function of the number of relocated people $C$ for the $6$
  recovery strategies in the $0.2 g$ PGA(a), $0.25 g$ PGA (b) and $0.3
  g$ PGA(c) earthquake scenarios. Each panel shows the absolute values
  of efficiency divided by the corresponding pre-shock efficiency
  $E_{cc}(0)$.}
\label{fig:eff_cit-cit}
\end{center}
\end{figure}

\begin{figure}
\begin{center}
\subfloat[][]{\includegraphics[trim=0cm 5cm 0cm 8cm, clip=true,
    angle=0,width=3in]{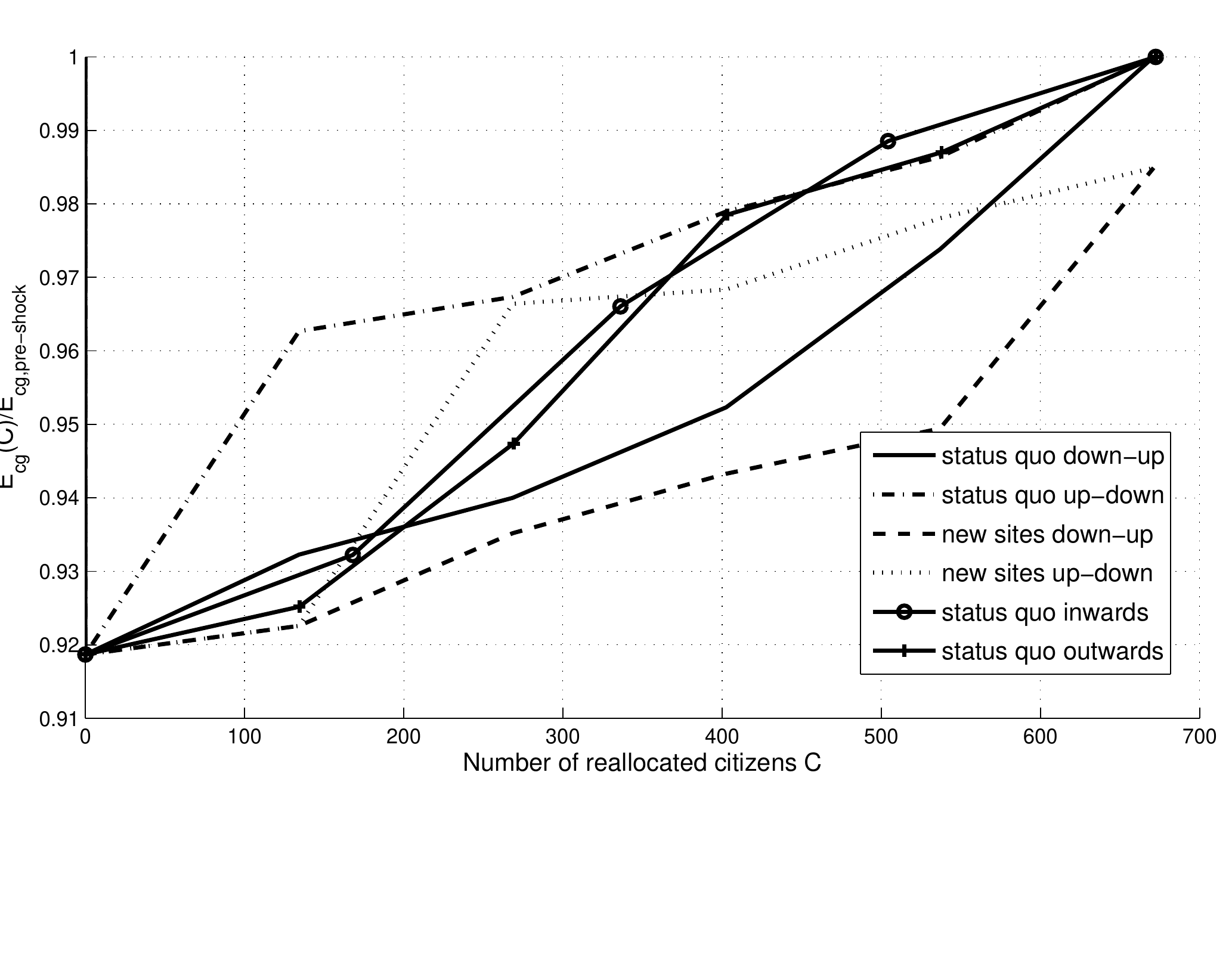}} \hfill
\subfloat[][]{\includegraphics[trim=0cm 5cm 0cm 8cm, clip=true,
    angle=0,width=3in]{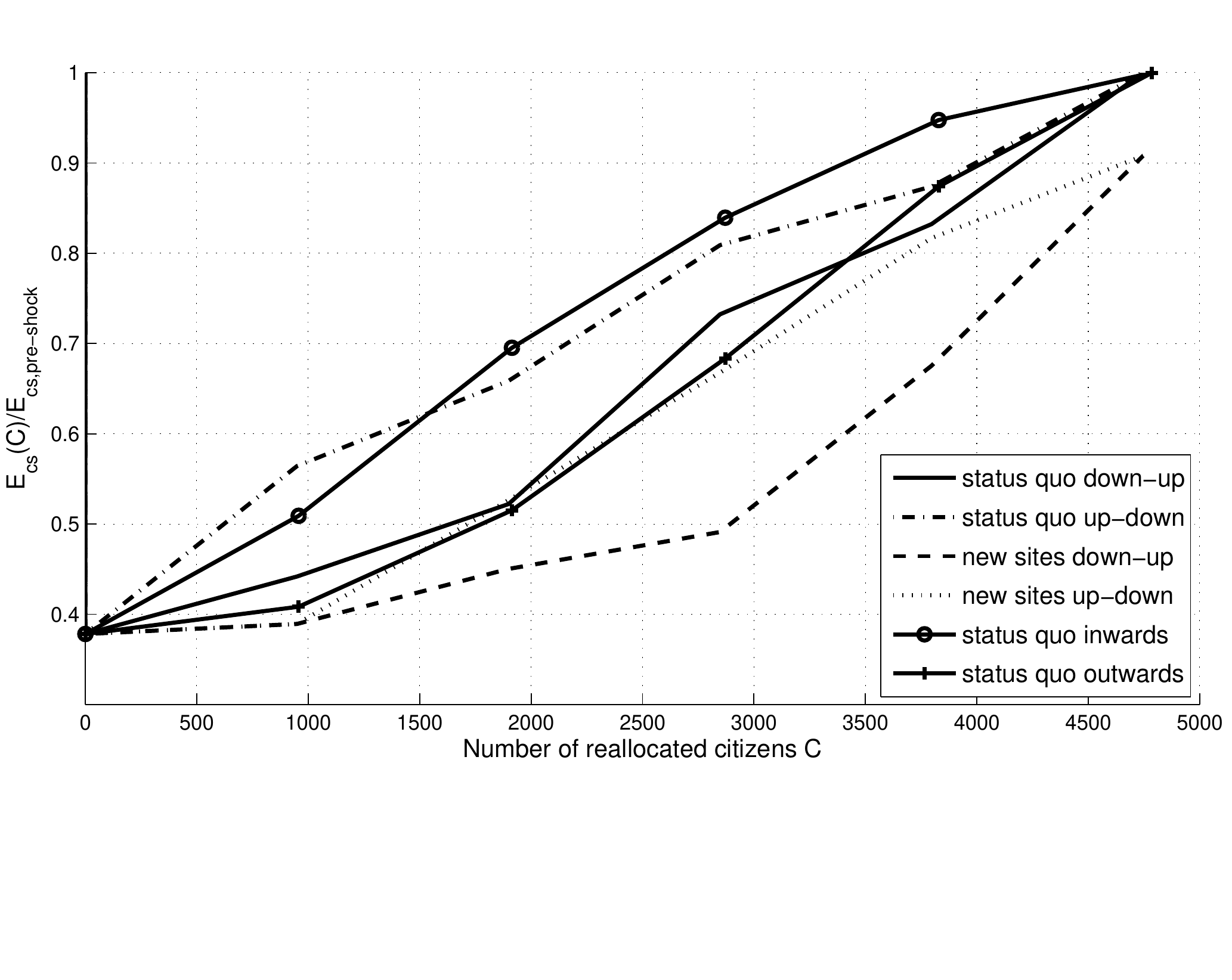}}
\\ \subfloat[][]{\includegraphics[trim=0cm 5cm 0cm 8cm, clip=true,
    angle=0,width=3in]{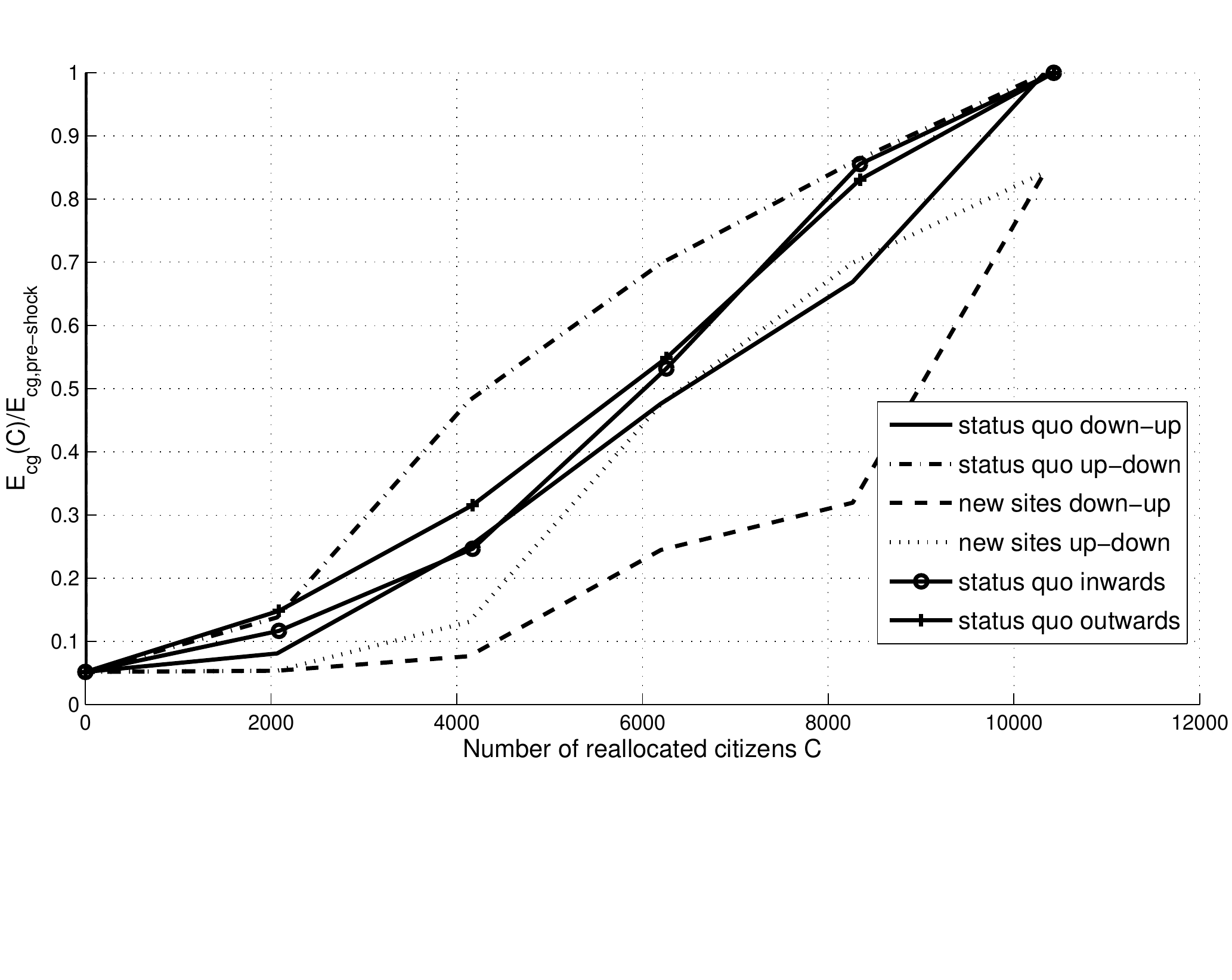}} \\
\caption{Values of $E_{cg}(C)$ for the $6$ recovery strategies in the
  $0.2 g$ PGA(a), $0.25 g$ PGA (b) and $0.3 g$ PGA(c) earthquake
  scenarios, as a function of $C$. As for
  Figure~\ref{fig:eff_cit-cit}, each panel reports the value of
  $E_{cg}$ at a certain reconstruction strategy divided by the
  efficiency of the original urban configuration.}
\label{fig:eff_cit-goods}
\end{center}
\end{figure}

\begin{figure}
\begin{center}
  \subfloat[][]{\includegraphics[trim=0cm 5cm 0cm 8cm, clip=true, angle=0,width=3in]{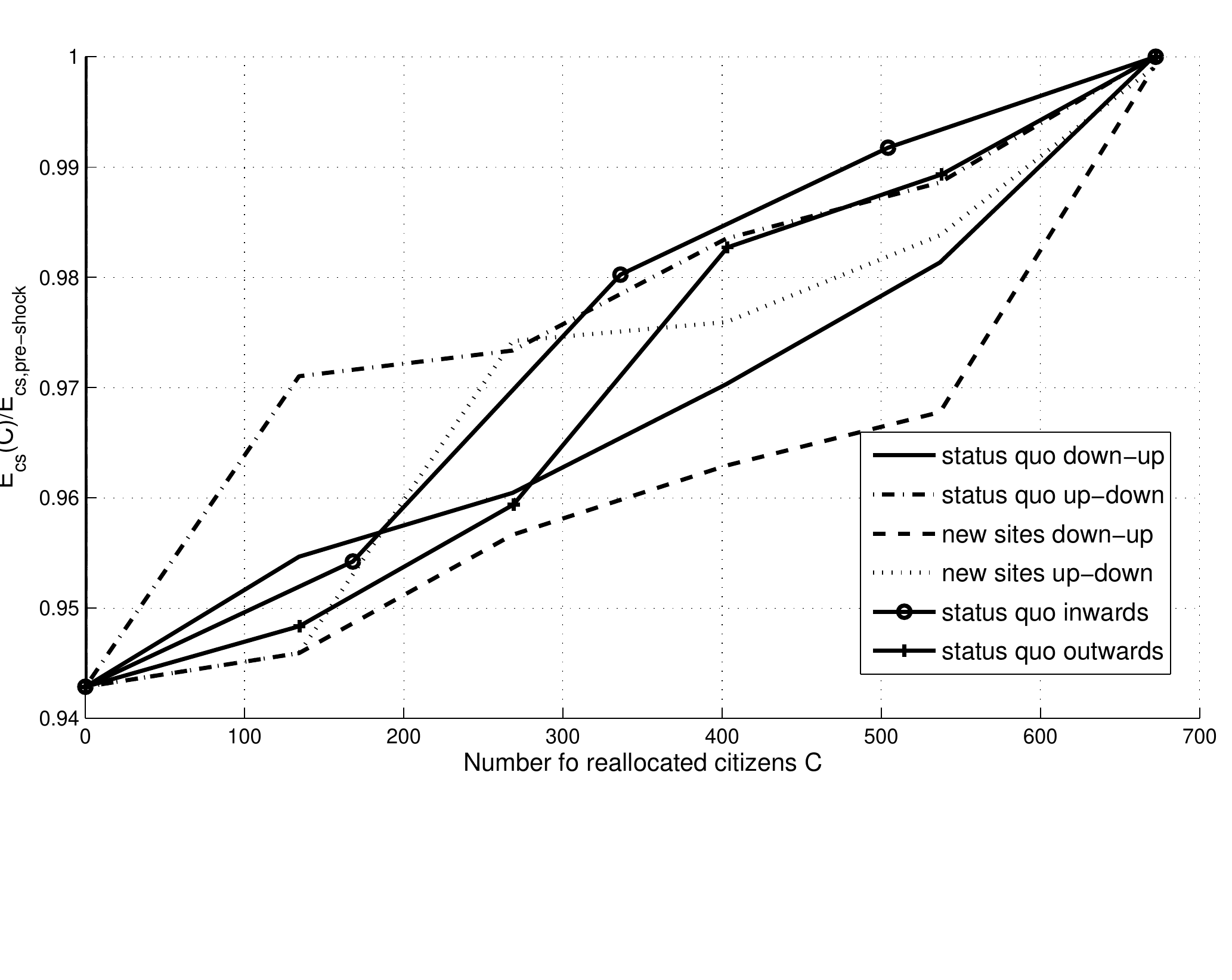}} \hfill
  \subfloat[][]{\includegraphics[trim=0cm 8cm 0cm 8cm, clip=true, angle=0,width=3in]{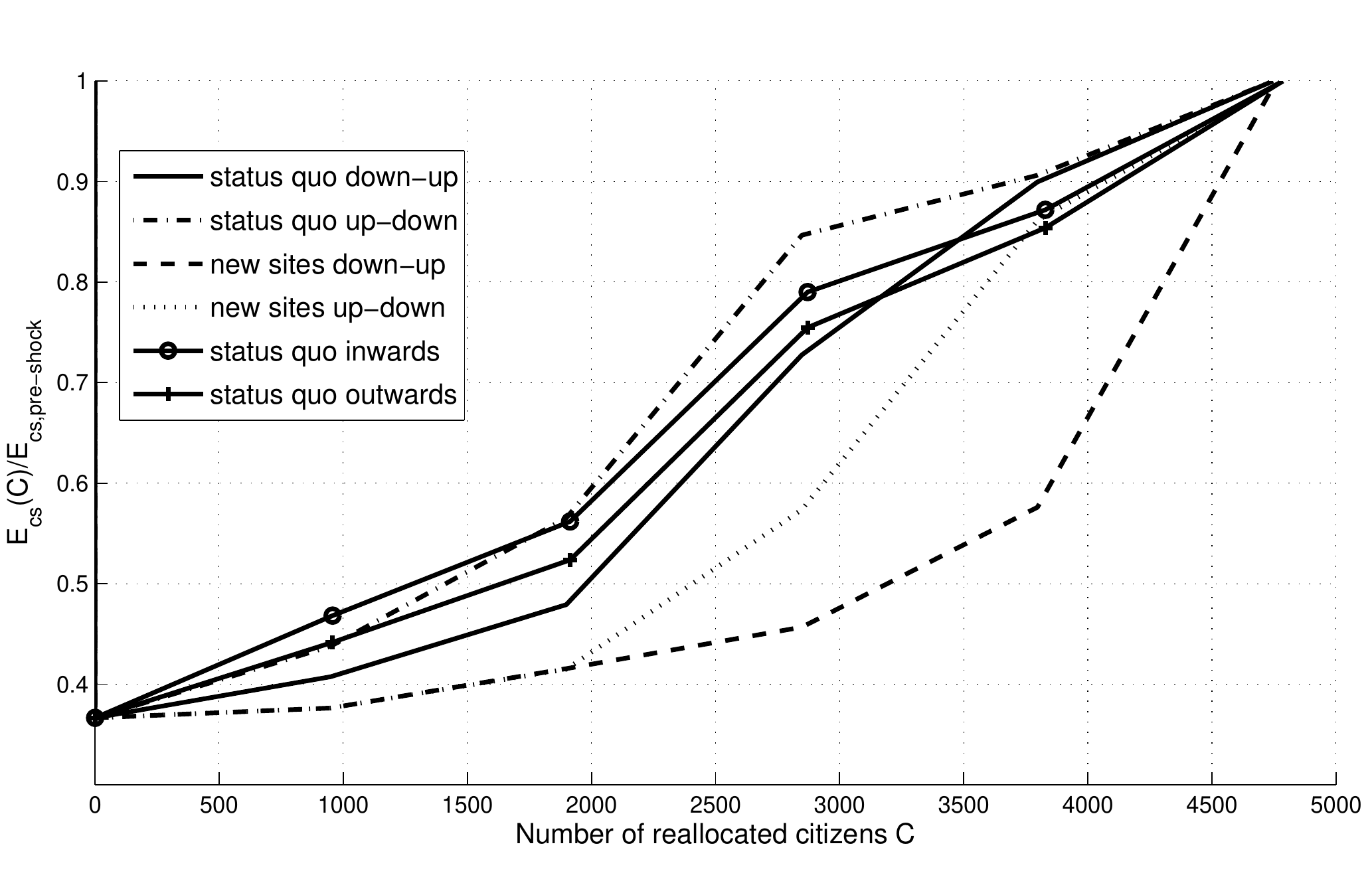}} \\
  \subfloat[][]{\includegraphics[trim=0cm 8cm 0cm 8cm, clip=true, angle=0,width=3in]{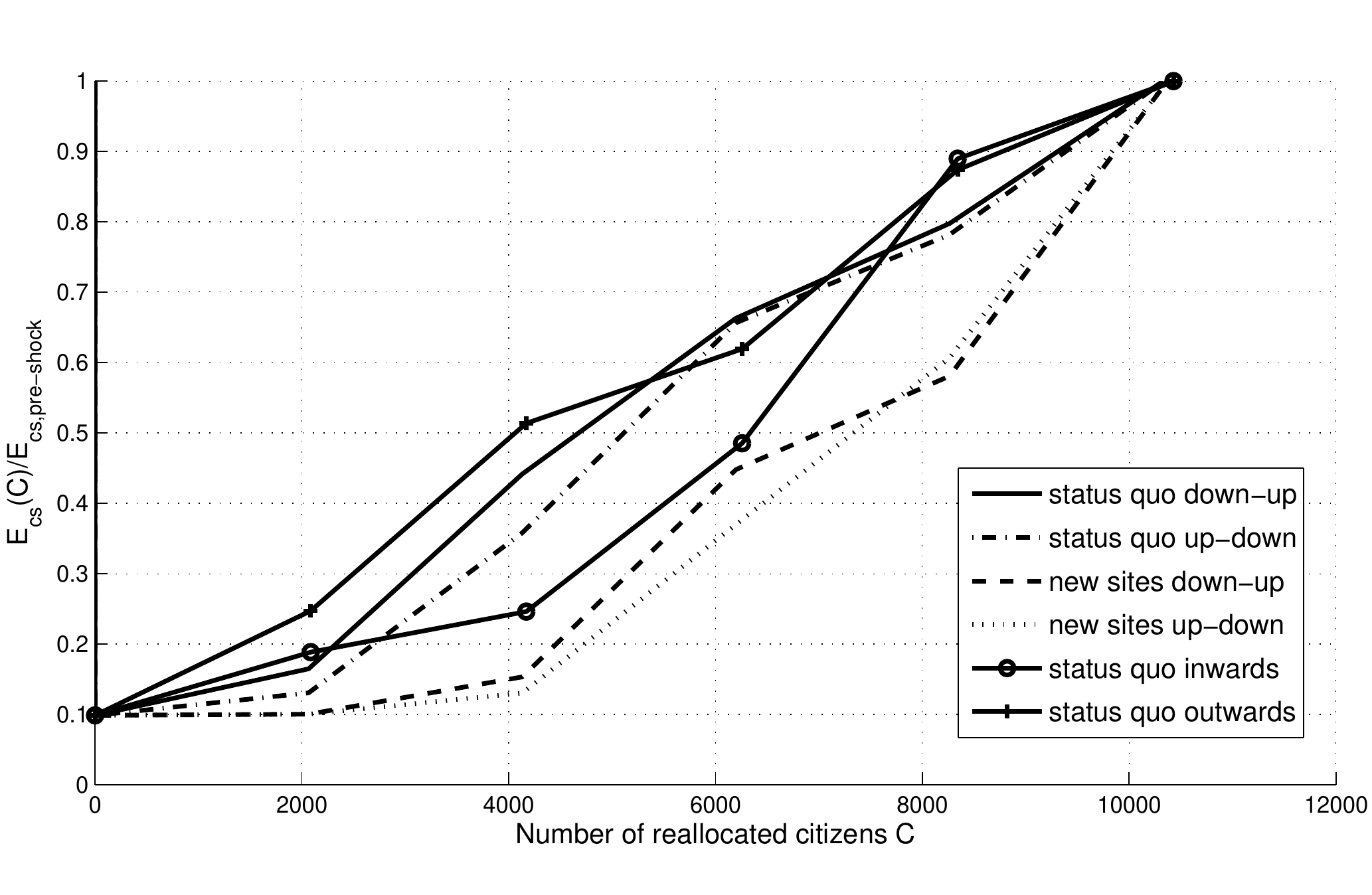}} \\
  \caption{$E_{cs}$ as a function of $C$ for the $6$ recovery strategies
    in the $0.2 g$ PGA(a), $0.25 g$ PGA (b) and $0.3 g$ PGA(c)
    earthquake scenarios.  The absolute values are normalized with the
    pre-shock efficiency $E_{cs}(0)$ }
  \label{fig:eff_cit-schools}
\end{center}
\end{figure}

For the comparative analysis of the six reconstruction strategies we
considered only three values of PGA, namely $0.2g$, $0.25g$ and
$0.3g$, which are consistent with the most likely eartquake intensity
in the region of Acerra\footnote{According to
  Figure~\ref{fig:hazard_acerra}, values of PGA larger than $0.3$ are
  very unlikely to occur.}. In Figure~\ref{fig:dam_scenarios} we
report a typical configuration of the simulated damaged network for
each of the three values of PGA. During recovery, we measured the
performance of the urban system by means of the efficiency of the
HSPN. The typical plots of $E_{cc}$, $E_{cg}$ and $E_{cs}$ of one
realisation corresponding to each value of PGA are reported in
Figure~\ref{fig:eff_cit-cit}, \ref{fig:eff_cit-goods} and
\ref{fig:eff_cit-schools}.

\begin{figure}
	\begin{center}
   \includegraphics[angle=0,width=3in]{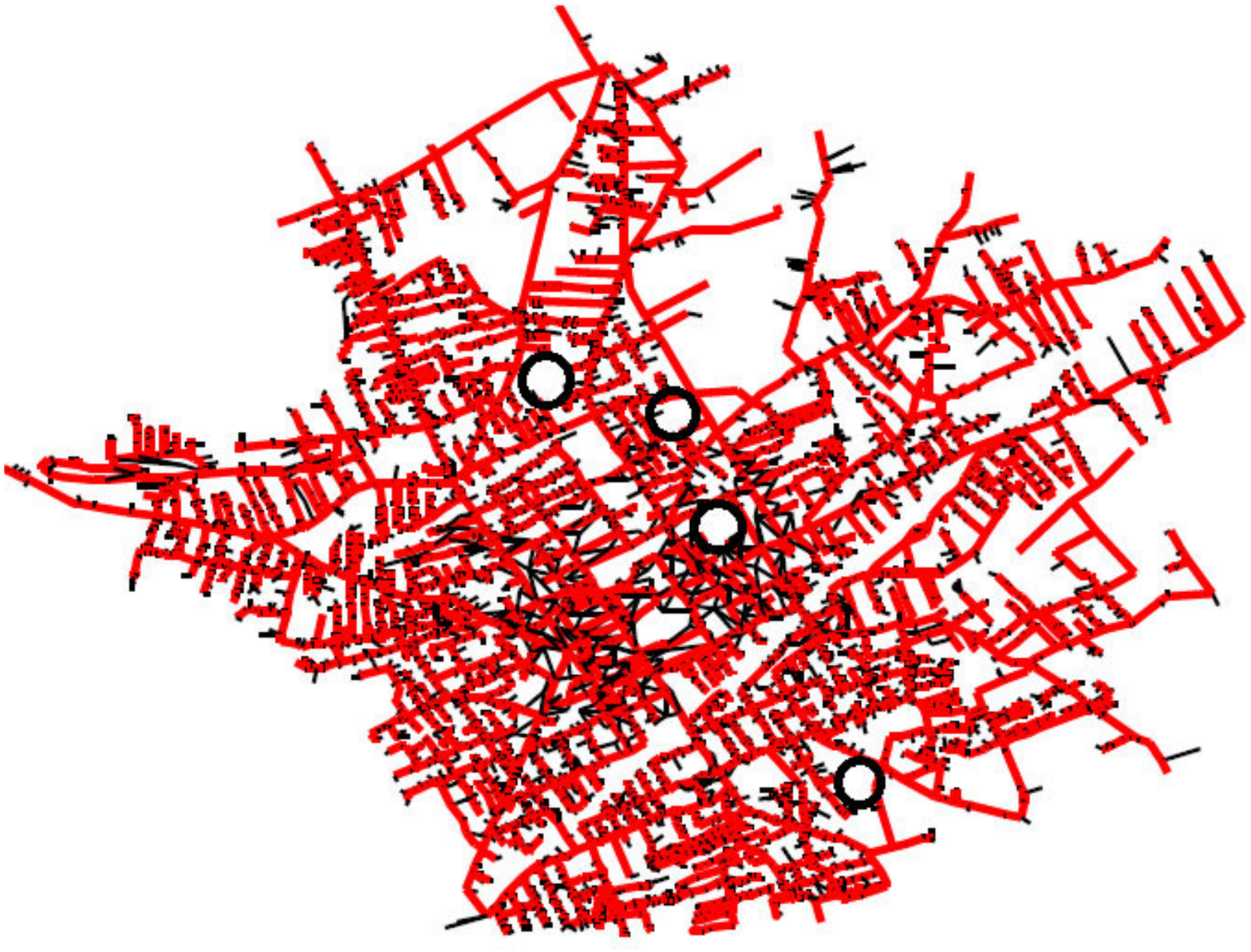}
	 \caption{The position of the new residential areas (circles) chosen
     for the \textit{new sites} recovery strategies.}
	\label{fig:empty}
	\end{center}
\end{figure}

\begin{table}[tp]%
  \caption {Values of $\mathcal{R}$ of the different reconstruction
    sttrategies, computed using definition~\eqref{eq:resilience} and
    averaged across several realisations. For each HSPN and each value
    of PGA, the $*$ indicates the most resilient reconstruction
    strategy, while the rows checked with $\dagger$ and $\ddagger$
    correspond to the two less resilient strategies.  }
\label{tab:resiliencies}\centering %
\footnotesize
\begin{tabular}{ccccc}
\hline
HSPN & PGA(g) & reconstruction strategy & $\langle \mathcal{R} \rangle$ & $\mathcal{\sigma_R}$	\\
\hline
cc &  0.2 & status quo outwards & 0.47391 & 0.08879 \\
* cc &  0.2 & status quo inwards & 0.55236 & 0.09428 \\
$\dagger$ cc &  0.2 & new sites up-down & 0.36180 & 0.09964 \\
$\ddagger$  cc &  0.2 & new sites down-up & 0.38107 & 0.08869 \\
cc &  0.2 & status quo up-down & 0.49051 & 0.08001 \\
cc &  0.2 & status quo down-up & 0.53831 & 0.08035 \\ \hline
cc & 0.25 & status quo outwards & 0.42057 & 0.08230 \\
* cc & 0.25 & status quo inwards & 0.56707 & 0.08151 \\
$\ddagger$   cc & 0.25 & new sites up-down & 0.33861 & 0.08539 \\
$\dagger$  cc & 0.25 & new sites down-up & 0.29228 & 0.09030\\
cc & 0.25 & status quo up-down & 0.48078  & 0.09865\\
cc & 0.25 & status quo down-up & 0.49351& 0.08014 \\ \hline
cc & 0.3 & status quo outwards & 0.43396 & 0.08190\\
cc & 0.3 & status quo inwards & 0.41964 & 0.08441\\
$\ddagger$  cc & 0.3 & new sites up-down & 0.28566 & 0.09984\\
$\dagger$  cc & 0.3 & new sites down-up & 0.27392 & 0.09762\\
cc & 0.3 & status quo up-down & 0.45144 & 0.08706\\
* cc & 0.3 & status quo down-up & 0.45853 & 0.09530\\
\hline
\hline
cg &  0.2 & status quo outwards & 0.49691 & 0.08089\\
cg &  0.2 & status quo inwards & 0.52764 & 0.09833\\
$\ddagger$   cg &  0.2 & new sites up-down & 0.48297 & 0.08193\\
$\dagger$ cg &  0.2 & new sites down-up & 0.26874 & 0.08987\\
*cg &  0.2 & status quo up-down & 0.63748 & 0.09920\\
cg &  0.2 & status quo down-up & 0.40420 & 0.09003\\ \hline
cg & 0.25 & status quo outwards & 0.41859 & 0.08798\\
* cg & 0.25 & status quo inwards & 0.57646 & 0.08102\\
$\ddagger$   cg & 0.25 & new sites up-down & 0.36596 & 0.08714\\
$\dagger$ cg & 0.25 & new sites down-up & 0.24454 & 0.08885\\
cg & 0.25 & status quo up-down & 0.53955 & 0.08797\\
cg & 0.25 & status quo down-up & 0.43379 & 0.09598\\ \hline
cg & 0.3 & status quo outwards & 0.44838 & 0.08595\\
cg & 0.3 & status quo inwards & 0.43294 & 0.08732\\
$\ddagger$   cg & 0.3 & new sites up-down & 0.32794 & 0.08493\\
$\dagger$ cg & 0.3 & new sites down-up & 0.18873 & 0.08525\\
* cg & 0.3 & status quo up-down & 0.51809 & 0.09438\\
cg & 0.3 & status quo down-up & 0.36390 & 0.09503\\
\hline \hline
cs &  0.2 & status quo outwards & 0.48476 & 0.08221\\
cs &  0.2 & status quo inwards & 0.56003 & 0.09898\\
$\ddagger$   cs &  0.2 & new sites up-down & 0.47966 & 0.08680\\
$\dagger$ cs &  0.2 & new sites down-up & 0.31011 & 0.09207\\
* cs &  0.2 & status quo up-down & 0.61563 & 0.08768\\
cs &  0.2 & status quo down-up & 0.42986 & 0.09178\\ \hline
cs & 0.25 & status quo outwards & 0.45244 & 0.09527\\
cs & 0.25 & status quo inwards & 0.49339 & 0.09459\\
$\ddagger$   cs & 0.25 & new sites up-down & 0.34012 & 0.08884\\
$\dagger$ cs & 0.25 & new sites down-up & 0.21159 & 0.09018\\
* cs & 0.25 & status quo up-down & 0.49917 & 0.09161\\
cs & 0.25 & status quo down-up & 0.43433 & 0.08902\\ \hline
* cs & 0.3 & status quo outwards & 0.50632 & 0.08900\\
cs & 0.3 & status quo inwards & 0.41086 & 0.09551\\
$\dagger$ cs & 0.3 & new sites up-down & 0.27636 & 0.08425\\
$\ddagger$   cs & 0.3 & new sites down-up & 0.30121 & 0.08320\\
cs & 0.3 & status quo up-down & 0.44357 & 0.08407\\
cs & 0.3 & status quo down-up & 0.46652 & 0.08789\\
\hline \hline
\end{tabular}
\end{table}

The values $\langle \mathcal{R} \rangle$, averaged over several
realisations of the three PGA scenarios, are reported in
Table~\ref{tab:resiliencies}, along with the corresponding standard
deviations $\mathcal{\sigma_R}$. The table suggests some remarks about
the ability of different reconstruction strategies to recover the
pristine urban efficiency.  First of all we observe that the
\textit{new sites} strategies provide quite slow efficiency recovery,
if compared with the \textit{status quo} strategies.  In particular,
in the \textit{new sites} strategies after the first step, consisting
in the reallocation of the $20\%$ of the inhabitants into new
buildings (the location of the four areas chosen for the simulated
construction of those four new buildings are shown in
Figure~\ref{fig:empty}), the efficiency values always show a
negligible increase. Furthermore, at the last step the efficiency
values are not totally recovered to their initial values.  This means
that the new configuration of the city, with $4$ new buildings used to
reallocate the $20\%$ of the displaced citizens, is less efficient
than its original configuration. Also, it was not possible to detect
any sensible difference between \textit{new sites up-down} and
\textit{new sites down-up}.

We notice that for earthquakes of PGA$=0.2g$ and PGA$=0.25g$, the
\textit{status quo} strategies (and in particular the \textit{status
  quo inward}) guarantee the fastest recovery of $E_{cc}$ among all
the reconstruction strategies considered.  This result can be
explained by the fact that, in this particular case study, the masonry
buildings, which are the most vulnerable to earthquakes, are mostly
placed in the centre of the city.  Hence, in the post-earthquake
configuration the center of the city will remain much more
disconnected, and in fact it is characterised by a considerably large
``hole'' at the center of the HSPN networks (as shown for instance in
Figure~\ref{fig:dam_scenarios}).  Thus, in the \textit{status quo
  outwards} strategy the first steps are not so efficient since the
building are restored onto a seriously damaged network. Conversely, in
the \textit{status quo inwards} strategy, the restored buildings are
progressively reinstalled onto a network which is also progressively
reconnected, making this strategy more resilient.

We would like to stress that the evaluation of the \textit{status quo
  inwards} strategy is of particular interest. In fact, in many old
cities the center consists mainly of ancient masonry structural
aggregates, and its reconstruction is particularly costly, both for
technical issues and for conservation constraints.  For these reasons,
reconstruction of damaged historical cities is often more rapid in the
suburbs than in the center.

\section{Conclusions}
\label{sec:conclusions}
In this paper we proposed a novel methodology to quantify the
resilience of complex social-physical urban systems against disasters.
This methodology aims to bridge the gap between the two classical
approaches to resilience: the engineering resilience, generally meant
as the capability of a system to recover its initial configuration
after a shock, and the ecosystem resilience, generally meant as the
capability of a system to recover its functionality even by reaching a
new configuration.
The procedure presented here is inspired by the idea that city
resilience should properly take into account its social components,
namely the citizens, which are the final users of the urban system as
a whole.  Our approach to quantify city resilience is based on the
efficiency of hybrid networks composed by citizens and urban
infrastructures.

In order to assess the capability of a city to recover its
functionality after a shock event, we compared the efficiency of the
corresponding hybrid networks before and after the shock event has
occurred and, as a case study, we considered the city of Acerra, for
which we simulated several earthquake scenarios and analysed
residential, goods and schools hybrid social-physical networks. We
also compared the ability of six different reconstruction strategies,
which differ from each other for the assignment of reconstruction
priorities, in restoring the pristine performance of the urban system.

While the main idea of the proposed methodology is to assess the
resilience of urban recovery, the quantification of the efficiency of
hybrid networks can be employed also for other purposes, e.g. to
compare the efficiency of different urban configurations or different
urban planning strategies (this would be a urban planning task), or to
design the reconstruction operations after an hazardous event. Thanks
to our approach, the best reconstruction strategy can be selected by
identifying the physical configuration which maximises the
performances of all the hybrid networks.

According to our metrics, the best strategy classes with respect to
residential reallocation are the \textit{down-up} and
\textit{inwards}.  In fact these strategies provide a faster response
in the immediate aftermath, securing and restoring many buildings in
the first reconstruction step.  In particular the \textit{status-quo}
strategies ensure the total recovery up to the original HSPN
efficiency.  The others strategies, allocating many citizens in a few
distant new areas produce longer average distance between people and,
consequently, a lower HSPN efficiency.  We can then infer that a good
reconstruction strategy should take care first of the restoration of
the bulk of a city. The analysis of the street network fragmentation
also underlines the weakness of the urban ecosystem of Acerra, in
which the overall connectivity of the HSPNs heavily depends on the
connectivity of a high-risk historical centre, consisting of ancient
masonry buildings which are indeed more prone to seismic risk.

We would like to stress that these results are strongly related with
the specific city configuration under study, and that extra care
should be taken while trying to generalize these results to other
specific cases.  We also notice that the methodology we proposed in
this work to quantify resilience is based only on the number of
reallocated citizens, and does not account for the availability of
financial resources, restoration rates or restoration prioritization,
also considering emergency management issues.  We believe that an
estimation of resilience based on the actual restoration time, i.e. on
the total time needed to restore the damaged buildings and to
reallocate all the citizens, would be possible by introducing in the
model additional information about financial resources and restoration
rates. We also notice that the availability of financial resources and
the prioritization strategies resulting from emergency management
issues are indeed a social-economic background input, not depending on
the adopted recovery strategy.

Once the financial aspect is introduced in the model, the methodology
could also be enriched and refined by considering the different
restoration costs associated to different building typologies. As a
matter of fact, is estimated that the cost of rehabilitating masonry
buildings in the historical centre, in Italy, is twice as larger than
the cost of rebuilding reinforced concrete structures in a similar
damage state, while the construction time in the first case is also
significantly higher than in the second. This difference in
restoration costs would affect the performance of the
\textit{outwards} strategies that start by rehabilitating the
historical centre, and would therefore be substantially less effective
than the \textit{inwards} strategies. This observation could become
even more important in the case of large scale disasters, where such
strategies would require to invest a large amount of resources in the
immediate aftermath of the disaster.

Further research activities on this topic are currently ongoing. In
particular, a research direction of interest is the quantification of
the impact of disasters on interconnected networks, by using some
recent theoretical results about the fragility of complex
interdependent
networks~\citep{Cohen2001,Cohen2000,Paul2004,Gao2011,Buldyrev2010}. Another
possible direction to explore is the usage of multiplex and
multi-layer networks, which have been recently proposed as a valuable
tool to model systems which consist of several different and
intrinsically interdependent systems~\citep{Morris2012,Brummitt2012}.

The authors are currently working to enrich and calibrate the measures
defined in this paper in order to propose a consistent set of
quantifiable efficiency measures to estimate the quality of life
perceived by the inhabitants of a urban systems, which could be used
also to measure the resilience of cities in an ecosystemic and
social-centric perspective.

\FloatBarrier


\end{document}